\documentclass[preprint,journal]{vgtc}       % preprint (journal style)

\ifpdf%                                % if we use pdflatex
  \pdfoutput=1\relax                   % create PDFs from pdfLaTeX
  \usepackage{graphicx}                % allow us to embed graphics files
  \DeclareGraphicsExtensions{.pdf,.png,.jpg,.jpeg} % for pdflatex we expect .pdf, .png, or .jpg files
\else%                                 % else we use pure latex
  \usepackage{graphicx}                % allow us to embed graphics files
  \DeclareGraphicsExtensions{.eps}     % for pure latex we expect eps files
\fi%

\graphicspath{{figures/}{pictures/}{images/}{./}} % where to search for the images

\usepackage{microtype}                 % use micro-typography (slightly more compact, better to read)
\PassOptionsToPackage{warn}{textcomp}  % to address font issues with \textrightarrow
\usepackage{textcomp}                  % use better special symbols
\usepackage{mathptmx}                  % use matching math font
\usepackage{times}                     % we use Times as the main font
\usepackage{cite}                      % needed to automatically sort the references
\usepackage{tabu}                      % only used for the table example
\usepackage{booktabs}                  % only used for the table example
\usepackage{multirow}
\usepackage{paralist}
\usepackage{balance}
\usepackage{tabularx}
\newcommand{\quan}{\textcolor{black}}
\onlineid{1525}

\vgtccategory{Research}
\vgtcpapertype{application/design study}

\title{RankAxis: Towards a Systematic Combination of Projection and Ranking in Multi-Attribute Data Exploration}

\author{Qiangqiang Liu, Yukun Ren, Zhihua Zhu, Dai Li, Xiaojuan Ma, and Quan Li}
\authorfooter{
\item
 Q. Liu was with ShanghaiTech and is now with Risk Department, Binance. This work was done when Qiangqiang Liu was a remote visiting student supervised by Quan Li at ShanghaiTech. E-mail: codi.l@binance.com.
\item
 Y. Ren, Z. Zhu and D. Li are with Corporate Development Group, Tencent. E-mail: rickyren,zhihuazhu,iamdaili@tencent.com.
\item
 X. Ma is with The Hong Kong University of Science and Technology. E-mail: mxj@cse.ust.hk.
 \item
 Q. Li is with School of Information Science and Technology, ShanghaiTech University. Quan Li is the corresponding author. E-mail: liquan@shanghaitech.edu.cn.
}

\shortauthortitle{Biv \MakeLowercase{\textit{et al.}}: Global Illumination for Fun and Profit}
\abstract{Projection and ranking are frequently used analysis techniques in multi-attribute data exploration. Both families of techniques help analysts with tasks such as identifying similarities between observations and determining ordered subgroups, and have shown good performances in multi-attribute data exploration. However, they often exhibit problems such as distorted projection layouts, obscure semantic interpretations, and non-intuitive effects produced by selecting a subset of (weighted) attributes. Moreover, few studies have attempted to combine projection and ranking into the same exploration space to complement each other's strengths and weaknesses. For this reason, we propose \textit{RankAxis}, a visual analytics system that systematically combines projection and ranking to facilitate the mutual interpretation of these two techniques and jointly support multi-attribute data exploration. A real-world case study, expert feedback, and a user study demonstrate the efficacy of \textit{RankAxis}.%
} % end of abstract

\keywords{Ranking, projection, multi-attribute data exploration}

\CCScatlist{ % not used in journal version
 \CCScat{K.6.1}{Management of Computing and Information Systems}%
{Project and People Management}{Life Cycle};
 \CCScat{K.7.m}{The Computing Profession}{Miscellaneous}{Ethics}
}

\teaser{
  \centering
\includegraphics[width=\linewidth]{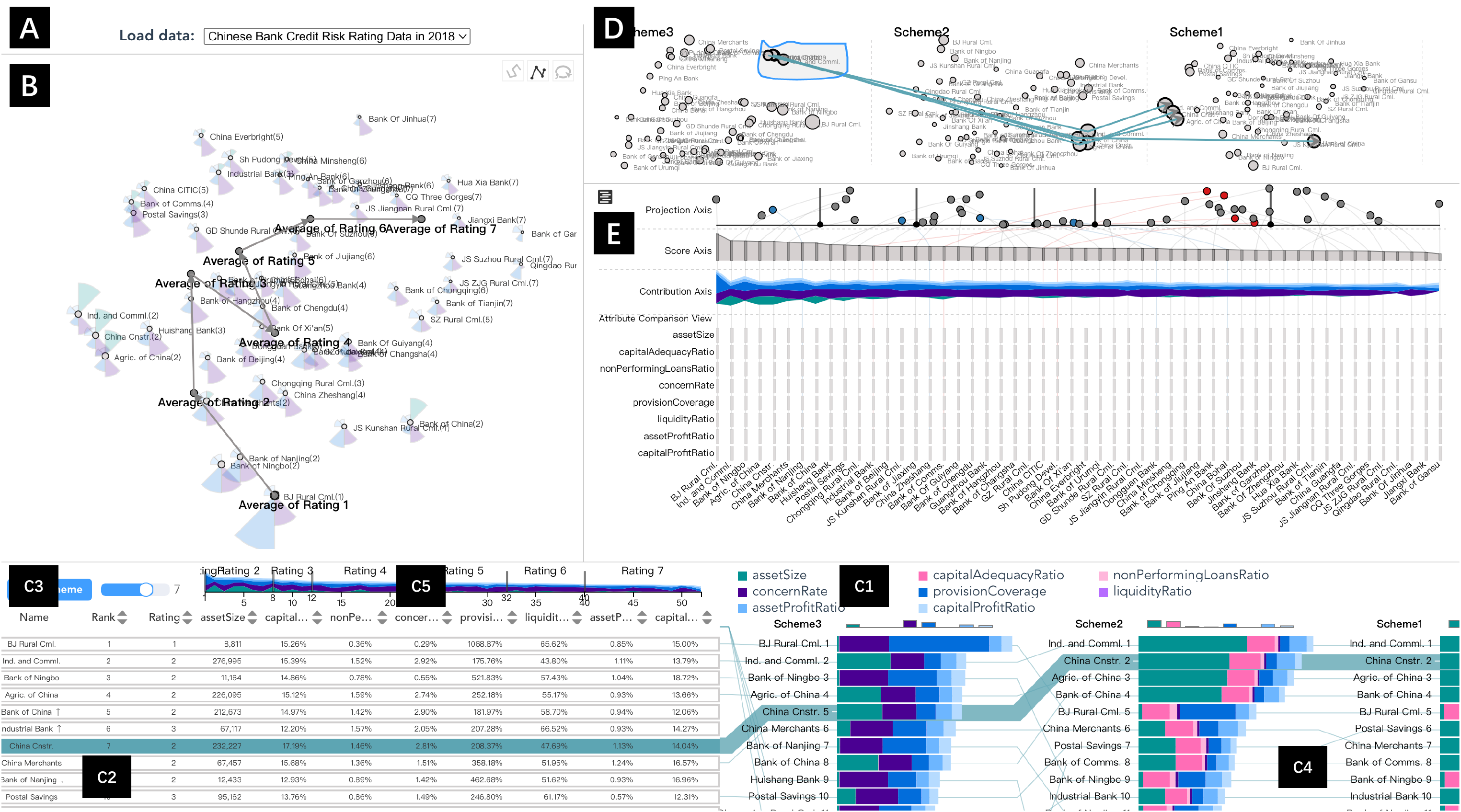}
\vspace{-6mm}
\caption{(A) The data loader facilitates data selection; (B) The interactive projection view shows the projection distribution and guides analysts to explore the projection layout and directional semantics; (C1 -- C5) The ranking tabular view summarizes the attribute contributions to the ranking and supports the deduction of the attribute weights based on user interaction, as well as compares different ranking schemes; (D) The comparative projection view analyzes the distribution of observations generated by different ranking schemes; (E) The ranking projection axis view compares the results of projection and ranking in the same context.}
\label{fig:teaser}
}

\vgtcinsertpkg

\begin{document}


\firstsection{Introduction}

\maketitle

\par Data analysts can choose from a variety of data exploration methods, each with its advantages and disadvantages in terms of visual interaction and performance ~\cite{grinstein2002information,munzner2014visualization}. For multi-attribute datasets presenting ``\textit{computational, design, and interaction tractability challenges}''~\cite{donoho2000high}, two common analysis methods are used, namely dimensionality reduction (i.e., projection) and multi-attribute ranking analysis~\cite{espadoto2019towards,li2018embeddingvis, zhao2017skylens}. The former is a visual abstraction for compressing a high-dimensional dataset into a low-dimensional space while maximizing the retention of attributes of the original structure and preserving pairwise relationships from the high-dimensional space in a low-dimensional projection. The latter reorganizes unordered datasets with multiple dimensions by computing ranking scores based on a single or multiple attribute(s), and is widely used in multi-criteria decision making (MCDM)~\cite{lai1994topsis,tzeng2011multiple}, allowing users to prioritize tasks or evaluate the relative performance.

\par Although projection and ranking are commonly used to elucidate large-scale, multivariate data, they have their inherent limitations \quan{in certain applications. For example, world-renowned rating agencies, such as Moody's, Standard \& Poor's and Fitch Rating, assess the financial and operational strength and risk resistance of countries, banks, securities firms, funds, bonds and public companies for the rating of financial institutions. Taking bank credit rating as an example. Each bank has several business-related metrics, such as asset size and delinquency rate, which can be viewed as a multi-attribute data item with numerical values. To solve the bank credit rating problem, dimension reduction and ranking can be applied. Specifically, in the projection, bank data items with multiple indicators are projected as observations in a low-dimensional space, which are then grouped by some unsupervised clustering algorithms. Some key metrics are selected to determine the order of clusters. In the ranking analysis, usually a set of bank attributes are selected and some MCDM schemes are used to rank all banks accordingly, and the generated ranked list is then divided into several segments, representing different bank ratings.}

\par \quan{However, letter-rated credits cannot be easily obtained through dimension reduction or ranking analysis alone before resolving the core issue in the showcased bank rating problem: ``\textit{ranking order occurs between institutions with different letter grades (similar to different rankings), but not between institutions with the same rating (similar to the same cluster)}''. First, projections employ a visual metaphor of ``proximity $\approx$ similarity'' that allows users to intuitively estimate the closeness between any two observations using a distance function in the projection space. This approach is a heuristic for interpreting data relationships. For example, users may want to identify ``clusters'' and ``outlier'' and their corresponding ratings in bank observations but specifying the boundaries of data clusters or outliers and their inclusion criteria may be not easy. Moreover, information is often lost during the projection process and thus it potentially produces a low-dimensional distribution that does not always accurately reflect the data relationships in the original high-dimensional space~\cite{wenskovitch2017towards,espadoto2019towards}. In other words, distant data items in the original space may end up being adjacent to each other once they are mapped to observations in the lower dimensional space. Second, although multi-attribute ranking can intuitively represent order, segmenting the ranking list into several semantic intervals can be difficult and uncertain. Analysts may want to know why a particular item in one interval is ranked lower or higher than others from another interval. However, determining how an attribute contributes to the ranking and how a change in one or several attribute(s) affects the ranking results is not a simple process~\cite{gratzl2013lineup}. In other words, setting or adjusting attribute weights can affect the ranking results of multiple attributes. In most cases, quantifying the importance of an attribute to ranking is challenging, and it is difficult for users to obtain a good set of weights to bring the final ranking in line with expectation.}

\par \quan{Inspired by the showcased rating problem, this study attempts to combine dimension reduction and multi-attribute ranking into the same exploration environment to obtain the best advantages of the two. Exploring the connections between them has several advantages. First, relative to projection in the showcased rating problem, multi-attribute ranking is more intuitive and controllable in directionality and can be potentially leveraged as the guidance for examining rating directions in the reduced-dimensional projection. Second, reduced-dimensional projection generates ``clusters'' and ``outliers'' that can assist analysts interactively and spatially in organizing observations on display, thereby making them more expressive. Moreover, those elements can potentially be employed in guiding further operations on a multi-attribute ranking list, such as segmenting the ranking list and selecting a subset of items or attributes for further investigation. Similarly, analysts can explain those ``clusters'' and ``outliers'' by aligning their evaluations with their findings adopted from the multi-attribute ranking results. For example, \textit{Explainers}~\cite{gleicher2013explainers} creates projection functions defined by their annotations, and the resulting derived dimensions represent the concepts defined by the user's example. Third, inconsistency may exist between the proximity among observations in the reduced-dimensional projection space and the ranking distance in the multi-attribute ranking result. For example, points of the same cluster in the projection have divergent rankings and vice versa, and they either or both may disagree with users' subjective perception. These inconsistencies may be due to the effects of selected attributes, or the different weights assigned to the attributes. Combining the results of reduced dimensional projection and ranking analysis for discovering and explaining these inconsistencies can help clarify the underlying data features.}

\par In this study, we introduce \textit{RankAxis}, an interactive visual analytics system that combines projection and multi-attribute ranking into the same exploration environment to facilitate the joint interactive interpretation of multi-attribute data. Specifically, we developed a \textit{Ranking Tabular View} that summarizes the contribution of attributes to the ranking, supports the derivation of attribute weights based on user interactions, and generates a ranking score for each data item. To help interpret the layout and orientation of the projection space, a \textit{Projection View} was designed to interact with the \textit{Ranking Tabular View} to produce a projection axis that guides the analyst to explicitly explore the layout and orientation of the projection. A \textit{Ranking Projection Axis View} is created to compare the ranking results, i.e. the ranking scores from the \textit{Ranking Tabular View}, with the projection axes generated in the \textit{Projection View} in the same context. This design also allows analysts to check for possible reasons of inconsistencies between ranking and projection results. We used a case study, a qualitative user study, and expert interviews to evaluate the efficacy of \textit{RankAxis}. Our main contributions are summarized below: (1) We systematically summarize the connection between projection and ranking and elicit design requirements from the literature and interviews with domain experts. (2) We design \textit{RankAxis}, an interactive visual analytics system that seamlessly combines projections and rankings in the same exploratory environment, helping analysts understand and interpret the results of projections and rankings. (3) We demonstrate the efficacy of our approach through one case study, a qualitative user study and expert interviews.

\section{Related Work}
\subsection{Visualizations for Multi-attribute Data}
\par Researchers have proposed many methods for analyzing and visualizing multi-attribute data, each with its own advantages and disadvantages~\cite{liu2016visualizing}. In particular, most methods propose specific solutions and trade-offs between data and scalability, and complexity and comprehensibility~\cite{blumenschein2018smartexplore}. For example, parallel coordinate plots~\cite{inselberg1985plane} drove the field forward and many subsequent improvements were proposed, e.g., highlighting density ~\cite{mcdonnell2008illustrative} and quality metrics~\cite{behrisch2018quality}, alleviating visual clutter~\cite{peng2004clutter,ellis2006plot} or by reordering dimensional axes to reveal certain patterns~\cite{dasgupta2010pargnostics}. Another popular approach, the scatter plot, uses orthogonal projections to analyze multi-attribute data and assess the usefulness of dimension combinations~\cite{tatu2009combining,albuquerque2010improving}. In addition to visualization, exploration techniques for navigation and user guidance have been proposed, e.g., \textit{LDSScanner}~\cite{xia2017ldsscanner} and Matrix/Tree~\cite{yuan2013dimension}. For example, Fernstad et al.~\cite{fernstad2013quality} proposed a quality metric-guided exploratory dimensionality reduction pipeline that enables users to interactively rank and weight variables based on the obtained quality metrics. \textit{SmartExplore}~\cite{blumenschein2018smartexplore} simplifies the identification and understanding of clusters, associations, and complex patterns in high-dimensional data through a table-based design that automatically selects and computes statistical metrics based on data attributes. In this work, we explore the connection between projections and rankings and focus on the mutual interpretation of their results, e.g., inconsistency and semantic orientation and layout.

%Previous research has shown that dimension reduction and clustering can be integrated into the same visualization~\cite{andrews2010space,endert2012semantics}, although they typically operate independently and in parallel, such that ``\textit{each technique supports some analytic component of the system without the influence of the other}''~\cite{wenskovitch2017towards}. 

\subsection{Dimension Reduction and Interactions}
\par Dimensionality reduction scales better in terms of sample size and dimensionality than other visualization methods, such as glyphs~\cite{yates2014visualizing} and parallel coordinates~\cite{inselberg1990parallel}. As a result, projections have become the preferred choice for exploring high-dimensional data and/or machine learning applications where individual properties of dimensionality are not as important. In particular, many projection techniques have been proposed~\cite{van2009dimensionality,sorzano2014survey,nonato2018multidimensional}, of which t-SNE~\cite{maaten2008visualizing} is arguably one of the best known and most adopted dimensionality reduction techniques. Espadoto et al.~\cite{espadoto2019towards} presented a quantitative survey on dimensionality reduction to choose the best technique for a particular usage context.

\par One challenge in exploring high-dimensional data using dimensional projections is that it is difficult for users to express their domain knowledge to ``\textit{steer the underlying data model}''~\cite{kwon2016axisketcher}, especially since they have little attribute-level knowledge. Wenskovitch et al.~\cite{wenskovitch2017towards} mentioned two types of interactions in the context of using dimensional projections, namely parametric interactions (PI) and observation-level interactions (OLI). The former refers to the direct manipulation of parameters to create a new projection. However, it can cause difficulties for novice or non-mathematically savvy analysts~\cite{wenskovitch2017towards}. On the other hand, OLI enables analysts to manipulate the observed data directly, insulating them from the complexity of the underlying mathematical model. For example, Li et al.~\cite{li2021semanticaxis} proposed \textit{SemanticAxis} that enables analysts to reconstruct projections by directly modifying attribute weights, which clearly falls under a PI. They also supported the creation of a semantic axis by selecting two sets of data observations. Nevertheless, their work only allows the analyst to examine clusters one by one (unipolar semantic axes) or two by two (bipolar semantic axes). Kim et al.~\cite{kim2015interaxis} proposed \textit{InterAxis} to properly interpret, define and change an axis in a user-driven manner. In particular, users can define and modify an axis by dragging data items to the $x$- or $y$-axis, and the system then computes a linear combination of data attributes and binds them to the axis so that the user can understand the axis and interact with it to adjust it accordingly. Subsequently, researchers proposed a technique for interpreting the user's drawing with an interactive, nonlinear axis mapping method called: \textit{AxisSketcher}~\cite{kwon2016axisketcher}, which enables the user to bring their domain knowledge by allowing interaction with data observations rather than attributes imposed in the visualization. \quan{Specifically, \textit{AxisSketcher} draws a curve in the projection according to users' perception, generating a high-dimensional curve as a new data axis, and the order of the data on the axis then reflects the user's knowledge. The data items are then projected onto the nonlinear axis, updating the scatter plot with it. The nonlinear axes established by \textit{AxisSketcher}  can better fit the user's conceptual model, but \textit{AxisSketcher}'s nonlinear axes distort the data locally, and this distortion affects the user's grasp and interpretation of the data facts. Our rating line is motivated by this work, with the rating line representing the perception of the data scores during the ranking process, with the addition of the concept of inconsistency.} For \textit{InterAxis}, users must be familiar with the dataset in order to select meaningful data items for further exploration, which ensures that the constructed linear axes are meaningful. For \textit{AxisSketcher}, users must have a basic understanding of the projection layout in order for the constructed non-linear axes to be meaningful. They both rely on the user's interaction intent; however, in many cases, users do not necessarily understand the data. As a result, many items or axes generated by interactions are not analytically meaningful, which can easily lead to unsatisfactory analysis. In this study, we propose a projection axis that is integrated with the ranking results, thus facilitating the understanding of the underlying data features.

\subsection{Multi-attribute Ranking Visualizations}
\par There are several standard visualization methods for multi-attribute ranking and are summarized~\cite{gratzl2013lineup}, including spreadsheet~\cite{fewdesigning}, point-based, line-based, and area-based techniques. In particular, the classic parallel coordinate diagrams ~\cite{inselberg1985plane}, slop graphs~\cite{tufte1985visual} and bump charts~\cite{tufte1990envisioning} are all part of line-based visual design. Tables with embedded bars~\cite{rao1994table}, multiple bars, and stacked bars~\cite{fewdesigning} belong to region-based techniques. In this study, we used stacked bars, a region-based technique, because it supports comparing the ranking of the same data item in different dimensions and with different ranking criteria~\cite{seo2004rank}.

\par Many spreadsheet systems for analyzing multi-attribute data, such as \textit{Microsoft Excel} or \textit{Numbers} for Mac, are primarily designed to generate, modify and present tabular data and are not designed for sorting analysis. Also, they largely do not support sorting based on attribute combinations, which would otherwise require high-level formulas to define the sort. Therefore, researchers have proposed a number of sorting systems that provide interactive interfaces. For example, \textit{ValueCharts}~\cite{carenini2004valuecharts} and \textit{LineUp}~\cite{gratzl2013lineup} allow analysts to create custom rankings with adjustable attribute weights by clicking and dragging attributes. Weng et al.~\cite{weng2018srvis} supported analytical tasks for ranking large-scale spatial alternatives, such as selecting the best location for a chain store. \textit{RankBooster}~\cite{puri2020rankbooster} goes further in understanding ranking predictions, i.e., what can be done to improve rankings. However, these systems require the user to specify attribute weights to produce rankings of data points, i.e., many assume that the user can conceptually quantify the understanding of how critical a particular attribute is to the ranking, which is not always easy or possible for the user to do.

\par To address this issue, researchers have studied what factors or weight sets lead to a given ranking. For example, \textit{Podium}~\cite{wall2017podium} allows users to drag rows in a table to rank data points based on their perception of the relative value of the data. \textit{WeightLifter}~\cite{pajer2016weightlifter} is an interactive visualization technique for weight-based MCDM that facilitates the exploration of weight spaces. Analysts can understand the sensitivity of decisions to changes in weights. In this study, similar to \textit{Podium} and \textit{WeightLifter}, we perceive attribute weights through user interaction. However, the weights derived from \textit{Podium} or \textit{WeightLifter} are not always applicable, as they do not guarantee satisfactory ranking results. In our work, we overcome the inconsistency problem by combining the results of projection and ranking into the same exploration setting to obtain the best results for both methods.

\section{Background and Requirement Analysis}
\subsection{Observational Study}
\par To better understand how multi-attribute data items are explored in practice and to further refine our design requirements, we conducted an observational study on a collaborative team of bank experts, including a bank rating specialist (E.1, male, age: $32$), a risk management specialist (E.2, male, age: $35$), a financial data analyst (E.3, female, age: $24$), and a bank credit specialist (E.4, male, age: $28$). Their daily work consists of digging into the impact of financial indicators of banks and other institutions over the years and developing criteria for the quantitative classification of institutions. As a rule, the experts use \textit{Excel} to periodically adjust the institution's rating based on their professional experience, combined with the bank's indicators over the years. Although manual methods can be used, they require a large amount of human resources to deal with numerous indicators simultaneously and are inconvenient for dynamic monitoring. Therefore, the experts tried to solve the bank credit rating problem by applying dimension reduction and ranking analysis. In the projection scheme, they first used some business intelligence (BI) tools (e.g., \textit{Tableau}) to project bank data items with multiple indicators as observations in a low-dimensional space. Then, the experts grouped these points using an unsupervised clustering algorithm. Finally, they specify some key metrics (e.g., average asset size) for each resultant cluster to determine the order of clustering. In the ranking analysis scheme, the experts first selected a set of bank attributes and then used specific \textit{MCDM} schemes such as \textit{Analytic Hierarchy Process (AHP)} and \textit{Statistical Product and Service Solutions (SPSS)} in \textit{Technique for Order Preference by Similarity to Ideal Solution (TOPSIS)} to rank all banks accordingly. Next, they divided the ranked list into segments based on the distribution of specific indicators. As a result, all banks in the higher segments outperformed the banks in the lower segments.

\par In addition to separate analyses using projection and ranking, experts have found that in some cases, the results of projection/clustering may not be consistent with the results produced by ranking. That is, bank items that are close to each other in one method are not necessarily close to each other in the other method. For the sake of simplicity, we formulate the inconsistency problem as follows. Suppose we have a set $R^n$ where each data item is a vector of $n$ dimensions, denoted as $x=\{a_1, a_2, ..., a_n\} \in R^n$ and assume that all attributes are positive, i.e., the higher the value the better. We use the same distance function in ranking and projection (e.g., Euclidean distance denoted by $f(x)$). Specifically, the ranking score of each $x_i$ is determined by the Euclidean distance between the original point and $x_i$, denoted by $f(x_i)$. The Euclidean distance between $x_i$ and $x_j$ in the projection space is denoted by $g(x_i, x_j)$. In the ranking, we decide which data item is ranked first based on the value of $f(x)$, i.e., if $f(x_i)>f(x_j)$, then $x_i$ should be ranked before $x_j$ and vice versa. This condition can be expressed as
\begin{equation}
\label{eq:1}
x_i \succ x_j \Leftrightarrow f(x_i) > f(x_j).
\end{equation}Suppose that in the projection, if $x_i$ and $x_j$ belong to the same ``cluster'' and $x_k$ belongs to another ``cluster'', the following equation holds according to the definition of distance in the projection,
\begin{equation}
\label{eq:2}
min(g(x_i, x_k), g(x_j, x_k)) > g(x_i, x_j).
\end{equation}
\par \textbf{Inconsistency} means that points of the same cluster in the projection have divergent rankings, and vice versa. If \autoref{eq:2} and either \autoref{eq:3} or \autoref{eq:4} hold, there is an inconsistency between the perceived distance in the projection and the ranking distance:
\begin{equation}
\label{eq:3}
f(x_i) > f(x_k) > f(x_j),
\end{equation}
  \vspace{-4mm}
\begin{equation}
\label{eq:4}
f(x_i) < f(x_k) < f(x_j).
\end{equation}
Both \autoref{eq:3} and \autoref{eq:4} indicate that the ranking of $x_k$ is in the middle of $x_i$ and $x_j$.
\par \textbf{Consistency} means that points in the same cluster in the projection are also close to each other in the ranking and vice versa. If \autoref{eq:2} and any of the following equation holds, the perceived distance in the projection and the ranking distance are consistent:
\begin{equation}
\label{eq:5}
min(f(x_i), f(x_j)) > f(x_k),
\end{equation}
  \vspace{-4mm}
\begin{equation}
\label{eq:6}
max(f(x_i), f(x_j)) < f(x_k).
\end{equation}
\autoref{eq:5} and \autoref{eq:6} indicate that both the ranking of $x_i$ and $x_j$ are lower or higher than that of $x_k$.

\par
\begin{figure}[h]
  \vspace{-3mm}
\centering % avoid the use of \begin{center}...\end{center} and use \centering instead (more compact)
\includegraphics[width=\linewidth]{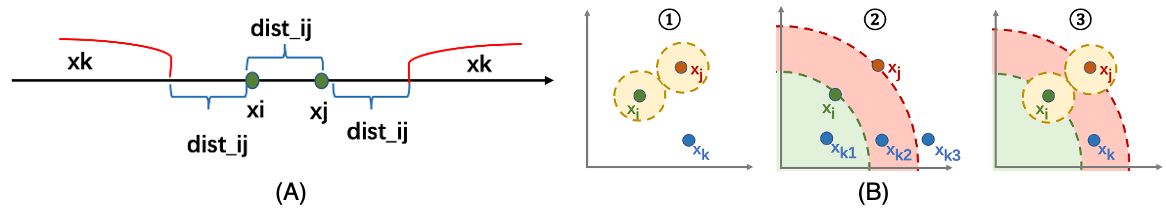}
\vspace{-8mm}
\caption{(A) When $n$ is $1$ and if \autoref{eq:2} holds, i.e., $x_i$ and $x_j$ are not in the same cluster with $x_k$, the distance between $x_i$ and $x_j$ in the projection must be greater than $dist_{ij}$ and $x_k$ can take the value in the red range. The distance in the projection and in the ranking are consistent; (B) Inconsistency occurs when $x_k$ lies in the circle between the green and red rings and not in the yellow circle.}
\label{fig:example}
\vspace{-3mm}
\end{figure}

 More specifically, if each data item has only one dimension (\autoref{fig:example}(A)), i.e., $n=1$, \autoref{eq:2} and \autoref{eq:5} or \autoref{eq:2} and \autoref{eq:6} hold. That is, the ranking distance is consistent with the projection distance. When the data item has two dimensions, i.e., $n$ is $2$, \autoref{eq:2} and \autoref{eq:5} hold or \autoref{eq:2} and \autoref{eq:6} hold. As shown in \autoref{fig:example}(B)(1), in the projection, the diameter of the yellow ring is the distance between $x_i$ and $x_j$. If \autoref{eq:2} holds, $x_k$ can be at any position in the non-yellow area in the projection, i.e., $x_{k1}$, $x_{k2}$, and $x_{k3}$. For ranking (\autoref{fig:example}(B)(2)), the distance between the point and the origin point indicates the ranking. Particularly, \autoref{eq:5} and \autoref{eq:6} hold for $x_{k1}$ and $x_{k3}$. \autoref{eq:4} holds for $x_{k2}$. In other words, if $x_k$ lies in the circle between the green and red rings and not in the yellow circle, \autoref{eq:2} and \autoref{eq:3} hold, which indicates that the distances in the projection do not match the ranking distances.

\subsection{Design Requirements}
\par \quan{To ensure that our approach was well suited to the tasks and requirements of the general field, we further interviewed all experts (E.1 – E.4) to identify their main concerns regarding bank ratings and potential barriers to efficient decision making. At the end of the interviews, the need for combining projection and ranking analysis in the same context emerged as a key theme in the feedback collected. While individual expectations for such an approach had different emphasis, specific design requirements were expressed across the board.}

\par \textbf{R.1 Connecting projections and rankings in a seamless context.} \quan{When experts explore bank projection results, they can observe several clusters of banks, but mapping different clusters to ratings can be a challenge. They encounter similar problems when interpreting the ranking results, as they rely on their expertise to divide the rankings into segments and subjectively use these segments as ratings. Therefore, experts wanted to put dimension reduction and ranking in the same context so that they could explore and compare them more effectively.}

\par \textbf{R.2 Automatic inference of attribute weights.} Traditionally, experts rank banks by listing and assigning weights that quantify the contribution of attributes. A frequent problem with this process is that they cannot effectively determine which specific attributes are important and to what extent, because they only have an overall understanding of the data. Both projection and ranking rely on distance functions to obtain the similarity of paired data items. However, the metrics that jointly calculate distances have different numerical scales and need to be normalized. For example, the metric of bank asset size is measured in trillions of dollars, while the default rate is a very small decimal. This situation creates a key problem: it is difficult to estimate the weight (contribution) of each attribute to the distance measure after normalization, even for experts who are already very familiar with the data. ``\textit{when we do a bank credit rating, we can know the approximate rating of the bank by looking at the specific values of certain attributes}'' (E.1). However, experts' subjective feelings about the importance of attributes are complex and difficult to quantify, so it is challenging to take this intuition and verify the appropriateness of the weights. Thus, simply asking experts to adjust the weights or using normalization can complicate the generation of results that meet the empirical expectations of domain experts. Therefore, they need to automatically infer the importance of attributes based on their perception of the data values.

\par \textbf{R.3 Guiding semantic exploration in projections.} Dimensional projections inevitably produce ``clusters'' and ``outliers'', and experts want to understand the distribution of observations in the projection space because they sometimes cannot distinguish between the boundaries of clusters and whether an observation is an outlier. That is, when interpreting the layout of the projection results, they want more semantic help to guide their exploration in the projection.

\par \textbf{R.4 Reveal any inconsistencies between projections and rankings.} As mentioned earlier, items projected together are not necessarily close in the ranking list, a situation that arouses the curiosity of experts because they regard ``proximity'' as ``similarity''. Similarly, they often confirm ranking results by observing whether the nearby neighbors of a data item in a ranked list are semantically related. Thus, identifying potential inconsistencies can help them better interpret the meaning of ``neighbors'' in projections and ranking results, and ultimately identify the underlying data characteristics that lead to inconsistencies.

\section{Back-End Engine}
\par
\begin{figure}[h]
   \vspace{-4mm}
 \centering % avoid the use of \begin{center}...\end{center} and use \centering instead (more compact)
 \includegraphics[width=\linewidth]{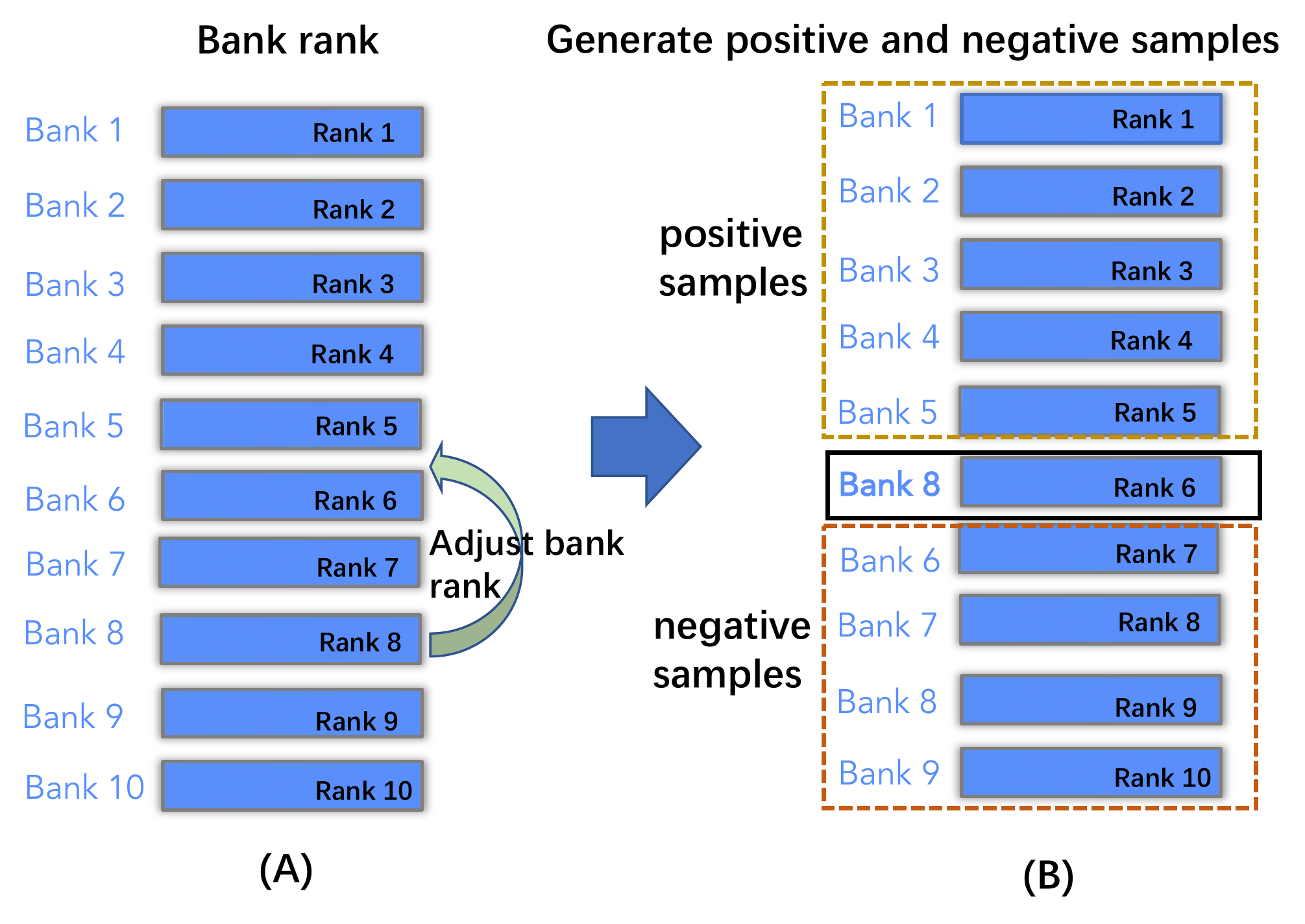}
     \vspace{-9mm}
 \caption{(A) The user can adjust the ranking of data items in the initial ranking result, then the system generates positive (the value of Bank 1-5 indicators minus the value of Bank 8 indicators) and negative samples (the value of Bank 6-9 minus the value of Bank 8 indicators), finally uses SVM to obtain the attribute weights.}
 \label{fig:svm}
   \vspace{-6mm}
\end{figure}

 In this section, we first describe how we use \textit{Ranking SVM} to automatically derive attribute weights~\cite{liu2021inspecting,joachims2002optimizing}. Then, we discuss how to obtain constraints from user-data interactions to train the Ranking SVM model (\autoref{fig:svm}(A)), and describe how to apply the weight vectors to produce a complete ranking of data items (steps 1 -- 3). Finally, we show how we transfer the rankings to the ratings (step 4).
\par \textbf{Step 1: Modeling Ranking SVM.} Inspired by \textit{Podium}~\cite{wall2017podium}, we use \textit{Ranking SVM} to derive attribute weights. \textit{Ranking SVM} applies the idea of optimizing the SVM hyperplane to the ranking problem with pairwise constraints. A finite set of data points $d_i$ and $d_j$ and a label is used to derive whether $d_i$ is better or not, instead of a complete set of data points with labels. The input to the \textit{Ranking SVM} involves a difference vector of data point pairs, e.g. $d_i - d_j$. Specifically, we transfer a pair $(d_i, d_j)$ and their relative ranks to a tuple based on the following statement: If $d_i$ is preferred, $d_i - d_j = 1$; otherwise, $d_i - d_j = -1$. The generated model can be used to predict which of the given pair of points is better. Nevertheless, the constraints derived from user interactions may be unsatisfying~\cite{joachims2002optimizing}. Therefore, we model all constraints as soft constraints rather than hard constraints to avoid vacuous results. Thus, user interactions can always produce a set of attribute weights that maximize the simulation of user constraints~\cite{liu2021inspecting}.

\par \textbf{Step 2: Deriving constraints.} We transfer ranking to a binary classification problem by using the linear separator of SVM. That is, we generate labeled data for \textit{Ranking SVM} by using the data items that the user has interacted with and dragging these items to a new location (\autoref{fig:svm}(B)). These items are the $k$ marked rows. Without loss of generality, the $k$ points $\{d_{l_1}, ..., d_{l_k}\}$ are indexed by $[l_1, ... , l_k]$. Then we create a combination of all pairs of difference vectors as training instances ~\cite{joachims2002optimizing}, i.e., for $i, j \in \{1...k\}$, where $i \ne j$, we derive a training tuple based on the above formula, i.e., each training instance is a pair of differences between rows $d_i$ and $d_j$, classified as $y = 1$ if $d_i$ is ranked higher than $d_j$, and $y = -1$ if $d_i$ is ranked lower than $d_j$. \quan{Similar to \textit{Podium}, we set $k = 6$ to ensure that the minimum training data amount for the attribute weight vector is derived after the experimental analysis.}

\par \textbf{Step 3: Calculating the ranking score.} After transforming the user interaction and learning the model, a weight vector $w$ is obtained for us to rank the data items. We compute the individual dot products of $w$ with each data item to generate a \textbf{rank score} as
$
r(d_i) = w \cdot d_i = \sum_{j=1}^mw_jd_{ij}.
$ with the highest one corresponding to the top rank.

\par \textbf{Step 4: Transfer ranking to rating.} We adapt an entropy discretization method to transfer rankings to ratings~\cite{fayyad1993multi}. We first sort ranking scores and consider each score as a split point, and then calculate the entropy of the left and right parts of each point. We consider the split point with the lowest entropy value to be the first split point. We repeat the above procedure until we have $n$ split points (we determine the value of $n$ for each dataset after the experimental results). We round the fraction of each data item to multiples of $n$. We denote the random variable of scores by $X$ and sort the scores of the data items as $(x_1, x_2, ..., x_n)$. The $P(x_i)$ denotes the probability of the fraction $x_i$. The entropy of $X$ is
$
H(X)=E[-logP(x_i)]=-\sum_{i=1}^NP(x_i)logP(x_i).
$ Suppose there are $k$ distinct scores among the ranking scores of all data items and $k<n$. We order $k$ scores as $(u_1, u_2, ... u_k)$, and these scores can be considered as $x_1, x_2, ..., x_n$ of consecutive values of breakpoints. Then, we select a point with the lowest entropy value from the candidate points. We repeat this process until we have $n-1$ breakpoints, forming $n$ ratings.

\section{Front-End Visualization}
\par \textit{RankAxis} combines projection and ranking to guide users in exploring multi-attribute data. In particular, we design a \textit{Ranking Tabular View} (\autoref{fig:teaser}(C1-C5)) that summarizes the attribute contribution to the ranking, deduces the attribute weight vector according to user interaction, and generates a ranking score. An \textit{Interactive Projection View} (\autoref{fig:teaser}(B)) shows the projection result and explicitly guides analysts to explore the projection layout and directional semantics by generating a projection axis. A \textit{Comparative Projection View} (\autoref{fig:teaser}(D)) evaluates the observation distributions generated by different ranking schemes. A \textit{Ranking Projection Axis View} (\autoref{fig:teaser}(E)) compares the ranking score from the \textit{Ranking Tabular View} with the projection axis generated in the \textit{Interactive Projection View} in the same context. The analyst can first interact with \textit{Ranking Tabular View} to obtain an initial ranking and rating result that guide the generation of projection axes in \textit{Interactive Projection View}, and then go to the \textit{Ranking Projection Axis View} to check for inconsistency between the ranking and projection results.

\subsection{Ranking Tabular View}
\par The ranking table view is designed with four purposes in mind. First, the raw multi-attribute data should be presented in a familiar Excel format to mimic the daily work of domain experts. Second, the contribution of each data attribute to the ranking should be visualized to facilitate the analyst's rating task. Third, the ranking can be adjusted interactively by moving a data item up or down based on the analyst's domain knowledge, thus automatically deriving attribute weights based on user interaction (\textbf{R.2}). Fourth, analysts should be able to compare the results of each adjustment. Different ranking schemes in terms of detailed attribute contributions for each data item should be provided, as well as the differences in ranking between the different ranking schemes, which is helpful to refine the previous rankings.

\par As shown in \autoref{fig:teaser}(C2), a table presents the raw data, showing the name, ranking, and associated attributes of the data items (\autoref{fig:teaser}(C1)). Analysts can perform a ``drag and drop'' operation to manually rank data items based on their perception and domain knowledge. In the table, the rankings that are adjusted higher are with up arrows, and down arrows indicates the opposite adjustment. By using the trained \textit{Ranking SVM}, the system derives a new weight vector that maximizes the user's data preferences. We calculate the attribute contributions by multiplying the obtained weights by the normalized attribute values. The sum of all attribute contributions (ranking score) is used to determine the order of the data items. The results of the attribute contributions are presented in \autoref{fig:teaser}(C5) and displayed at the top of the data table. Specifically, the position of the black lines represents the boundary between two ratings, while the colored areas indicate the attribute contributions. The region chart ranks all data items from left to right according to their ranking. Based on the ranking scores, we divide the rankings into several ratings according to the information entropy-based discretization algorithm mentioned earlier. For example, in \autoref{fig:teaser}(C5), data items are divided into five ratings separated by four automatically generated black lines. We also let the user interactively adjust the number of ranks by moving the slider next to the ``save scheme'' button (\autoref{fig:teaser}(C3)). If analysts is satisfied with the current ranking scheme, they can click on the button and the current ranking scheme will be added to \autoref{fig:teaser}(C4). Inspired by \textit{Lineup}~\cite{gratzl2013lineup}, each ranking is depicted as a separate column in \autoref{fig:teaser}(C4). These columns use bars of different lengths to represent attribute contributions. To compare ranking schemes, we arrange them horizontally, using lines connecting the same data items in different ranking schemes. When a specific data item is selected, a thick blue line connects all the same data items in turn. The bar chart at the top of each column (\autoref{fig:teaser}(C4)) shows the weights of each data attribute.

\begin{figure}[h]
     \vspace{-4mm}
 \centering
 \includegraphics[width=\linewidth]{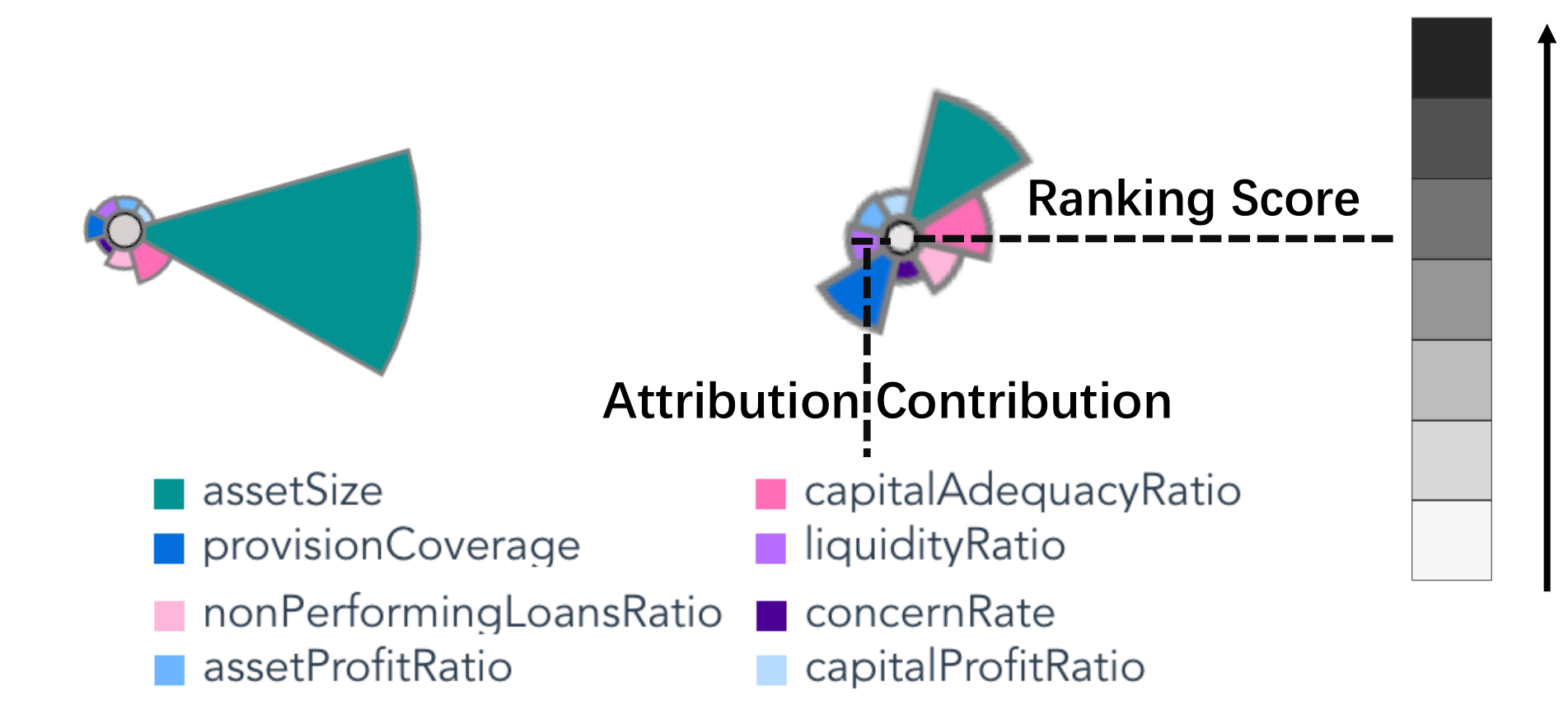}
     \vspace{-8mm}
 \caption{Design of observations in the interactive projection view.}
 \label{fig:projectiondesign}
       \vspace{-4mm}
\end{figure}

\begin{figure*}[h]
 \centering % avoid the use of \begin{center}...\end{center} and use \centering instead (more compact)
 \includegraphics[width=\linewidth]{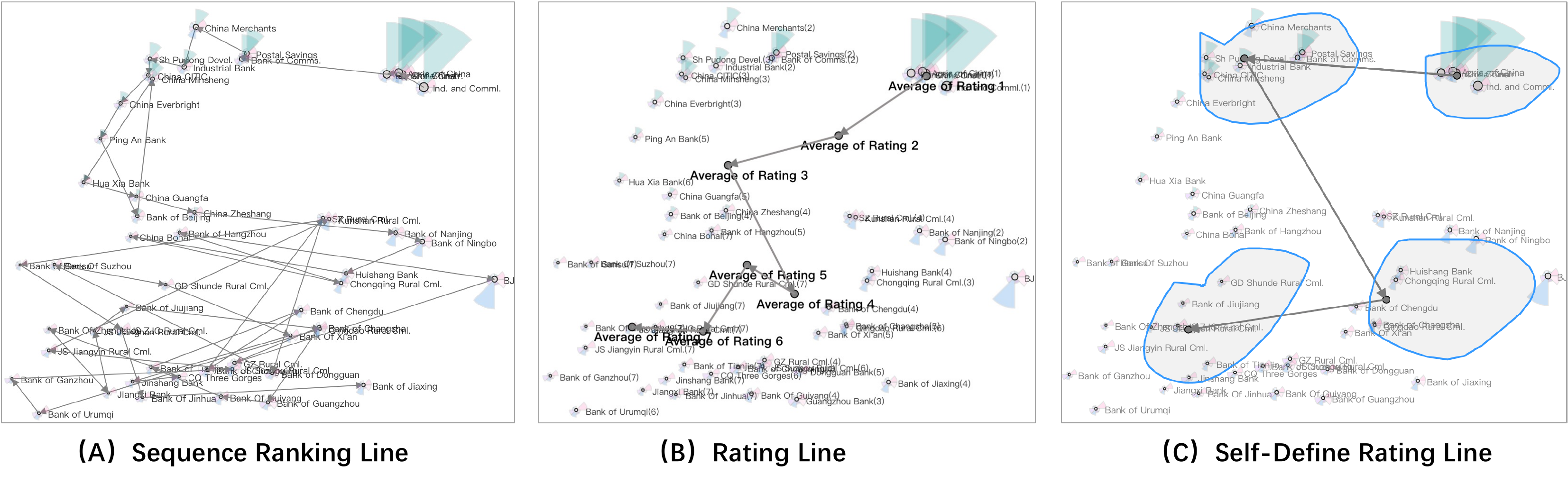}
     \vspace{-8mm}
 \caption{(A) Sequence ranking line connects each observation in interactive projection view in the order from the best to the worst according to the ranking results in ranking tabular view; (B) Rating line connects the average observation of each rating according to the order of ratings; (C) Self-defined rating line allows users to lasso a series of areas according to their semantic understanding and connects the average observations of all areas according to the order selected by the users.}
 \label{fig:rankingline}
      \vspace{-6mm}
\end{figure*}

\subsection{Projection View}
\par The projection view uses classical dimensionality reduction techniques such as \textit{t-SNE} to create low-dimensional projections and preserve local similarity to express neighborhood structure~\cite{2018Interactive,li2018embeddingvis}. Other techniques such as \textit{PCA}, \textit{MDS} and \textit{UMAP}~\cite{Mcinnes2018UMAP} can also be integrated. We use the same set of weights to normalize the values of the attributes used in the ranked table view to obtain a two-dimensional projection. Specifically, the four projection views are depicted in the \autoref{fig:teaser}. \autoref{fig:teaser}(B) shows the projection using the latest attribute weight vector. \autoref{fig:teaser}(D) shows the projection corresponding to the attribute weight vectors from the first three ranking schemes (\textbf{R.1}). We first introduce the interactive projection view and then the projection view for comparison.

\subsubsection{Interactive Projection View}
\par In the interactive projection view, there are two main components. First, as shown in \autoref{fig:projectiondesign}, observations on the interactive projection view are coded by a coxcomb digram~\cite {kammer2020glyphboard} that shows the distribution of attribute contributions. The color of the dot in the middle of the glyph encodes the ranking score: the higher the ranking score, the darker the color. The size of each pie encodes the corresponding attribute contribution. We did not choose the classic star glyph to encode attribute values because the lines in the star glyph are difficult to detect when the color saturation is low and the glyph is small~\cite{zhao2017skylens}. A potential drawback of this design is visual clutter, which is a common problem for many reduced-dimensional based visualizations. To mitigate this problem, we first reduced the opacity of the glyphs so that individual glyphs could be observed. When hovering over a glyph, that glyph is zoomed in and displayed in the foreground. In addition, we support panning and semantic zooming to focus on specific areas of the glyph. Second, the interactive projection view can generate a ranking line that explicitly guides the analyst to explore the projection layout and orientation (\textbf{R.3}). We define a ranking line that connects certain sampling points according to their rankings. To reflect the ranking results in the ranking table view, we propose the following three methods to generate ranking lines in the interactive projection view. Other methods, such as clustering based on projection quality or pressure~\cite{coimbra2021analyzing} can also be integrated to recommend initial clustering and avoid potential misinterpretations.

\par \textbf{\textit{Sequence Ranking Line.}} Based on the ranking score of each data item in the ranking tabular view, a line with arrows connects the corresponding observations in the interactive projection view (\autoref{fig:rankingline}(A)). We observe that the connected layout may show a trending order, e.g., an ordered ranking line in one direction, or a ``zigzag'' ranking line with backward and forward correspondence.

\par \textbf{\textit{Rating Line.}} Although a sequence ranking line may indicate an ordered layout, it strings all projection observations and may inevitably introduce visual clutter. For example, if ``zigzags'' occur frequently, then ranking lines do not adequately reflect the sequential semantic information contained in the projection space. It may also be difficult for users to keep track of the order of ranked data items. To alleviate this problem, we first partition the ranking results to obtain a subset of sequences with sequential different ratings. Then, we generate the average of the data items for each rating as the center of the rating in the projected view. These newly generated observations are highlighted in red and linked according to the order of the ratings (\autoref{fig:rankingline}(B)). That is, the generation of rating lines to link ``average observations'' in the interactive projection view can be considered as a ``resampling'' of the observations in the sequential ranking line, better reflecting the sequential semantic information contained in the projection.

\par \textbf{\textit{Self-defined Rating Line.}} The first two methods draw lines based on rating results, but ignore users with extensive domain knowledge. For example, with respect to bank rating questions, joint-stock commercial banks generally outperform private banks. In the interactive projection view, analysts may be inclined to conclude that the regions where the joint-stock banks are located are likely to be the better performers overall because they use the visual metaphor of ``proximity $\approx$ similarity''. Therefore, analysts can perform customized interactive operations to generate ranking lines based on their judgment of the data. As shown in \autoref{fig:rankingline}(C), analysts can lasso a region, and then the system automatically calculates the average of all observations in that region and joins all ``average observations'' generated from the lassoed regions in the order of user interaction to form a user-defined rating line.

\begin{figure}[h]
   \vspace{-3mm}
 \centering % avoid the use of \begin{center}...\end{center} and use \centering instead (more compact)
 \includegraphics[width=\linewidth]{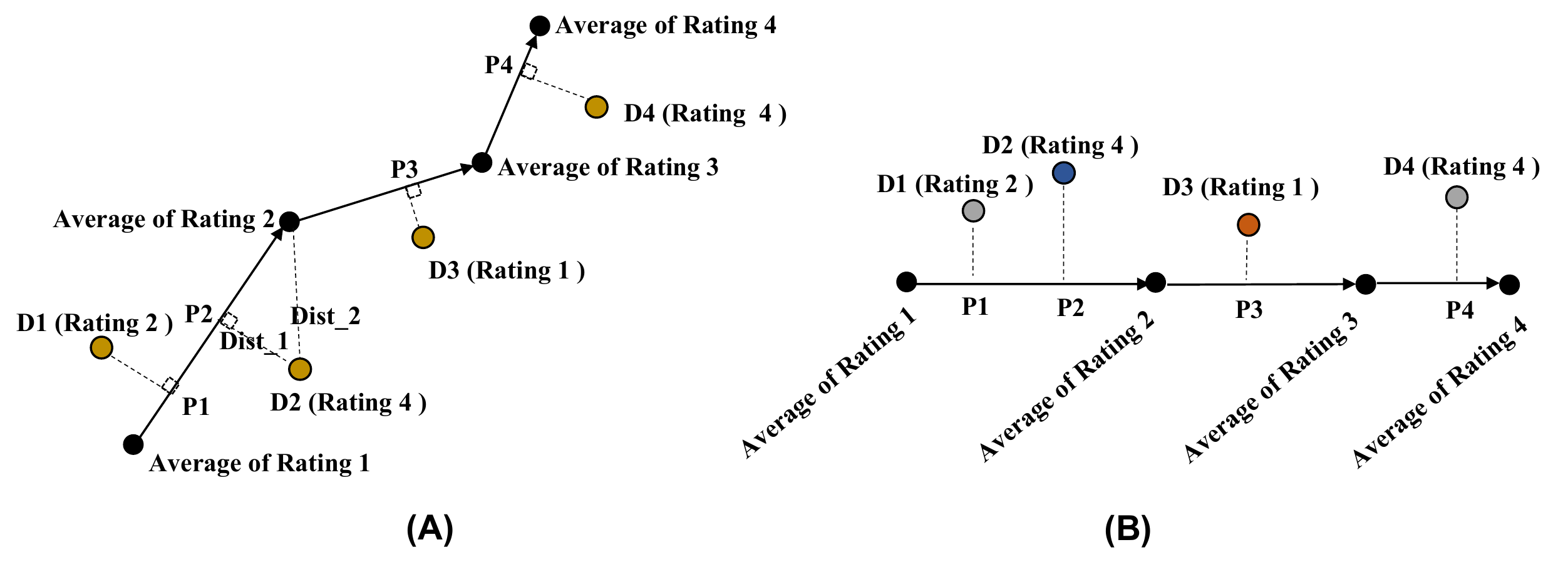}
     \vspace{-8mm}
 \caption{(A) The observations in the projection view (yellow dots) are projected on the ranking line (solid line with arrows); (B) the ranking line is expanded to a straight line to form a projection axis.}
 \label{fig:projectionaxis}
   \vspace{-3mm}
\end{figure}

\begin{figure*}[h]
 \centering % avoid the use of \begin{center}...\end{center} and use \centering instead (more compact)
 \includegraphics[width=\linewidth]{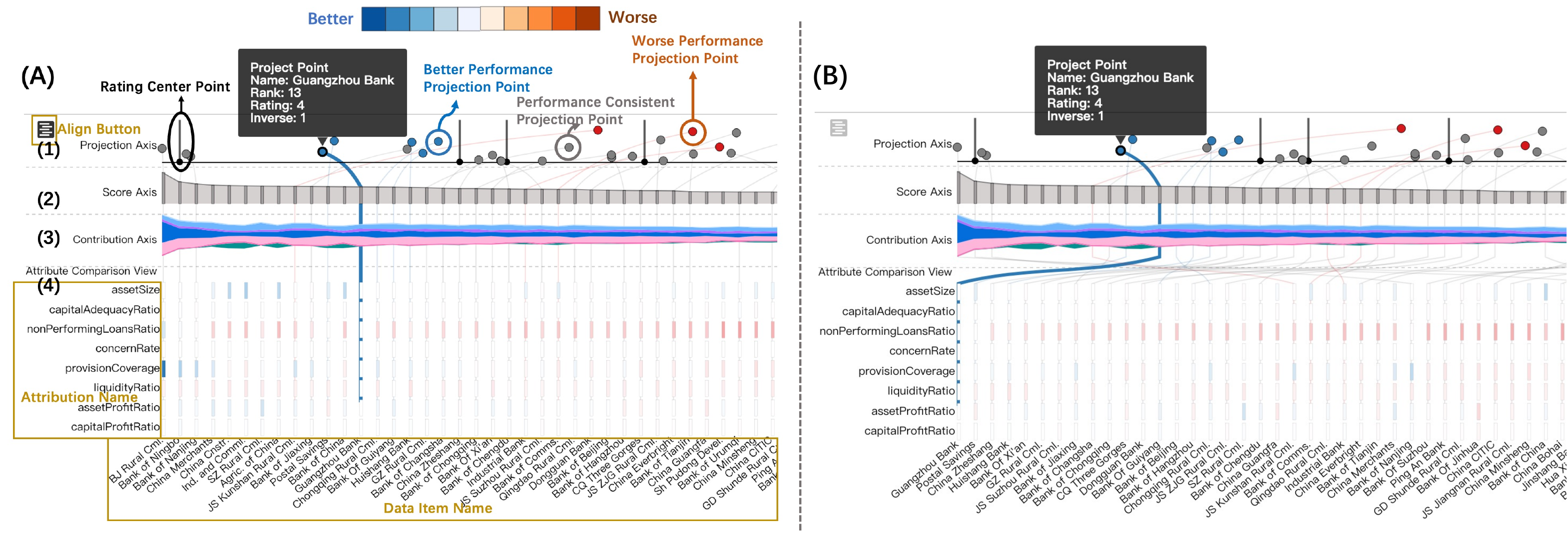}
     \vspace{-8mm}
 \caption{(A) Ranking Projection Axis View consists of four parts: (1) projection axis; (2) score axis; (3) contribution axis; and (4) attribute comparison view. (B) Resort the banks by clicking on the ``align'' button for attribute similarity comparison.}
      \vspace{-6mm}
 \label{fig:rankingprojectionaxis}
\end{figure*}

\subsubsection{Generating Projection Axis}
\par We generate the projection axis with the help of the previous ranking lines. As shown in \autoref{fig:projectionaxis}(A), the solid line with arrows is the ranking line and the yellow dots ($D1$, $D2$, $D3$ and $D4$) represent data observations in the interactive projection view. For example, in the case of the rating line, each turning point on the ranking line represents a rank, e.g.: \textit{Average of Rating 1} The $D2(Rating 4)$ indicates that the data item $D2$ has a rating of $Rating 4$. Then, we project the yellow observations onto the ranking line and calculate the shortest distance from the projected observations to the ranking line. For example, with respect to $D2$, the shortest distance to the ranking line segment ($Rating1$, $Rating2$) is $Dist_1$, while the ranking line segment ($Rating2$, $Rating3$) is $Dist_2$. Note that $Dist_1 < Dist_2$. Therefore, $D2$ is projected onto the ($Rating1$, $Rating2$) segment labeled $D2 /in (Rating1, Rating2)$. We expand the ranking line to obtain the projection axis (\autoref{fig:projectionaxis}(B)). The height of each data point is equal to the distance between the observation in the projection and its nearest ranking line, allowing us to encode the probability that a data point belongs to a certain rank. For example, $Dist_1$ represents the probability that $D2 \in (Rating1, Rating2)$, such that the shorter the distance, the higher the probability that $D2 \in (Rating 1, Rating 2)$. According to \autoref{fig:projectionaxis}(B), we have $D2 (Rating4) \in (Rating1, Rating2)$ and $D3 (Rating1) \in (Rating2, Rating3)$. That is, in the ranking results, $D2$ belongs to $Rating4$, but its position in the projection falls between $Rating1$ and $Rating2$, thus indicating that its semantic classification in the projection is improved. Therefore, we mark the color of $D2$ as blue and the color of $D3$ as orange red (blue indicates improved semantic classification in the projection and orange red indicates the opposite). The visual design of the projection axes is described in Section 5.3.

\subsubsection{Comparative Projection View}
\par \autoref{fig:teaser}(D) connects \textit{t-SNE} projections according to the attributes and weights of the different ranking schemes (in this study, we set the number of projection spaces to $3$). The analyst can lasso the observations on any space and all the same observations will be connected by curves. Regarding design alternatives, we initially used animations to track the shift of observations between the two rankings. However, our experts reported that tracking an observation requires a mental comparison, an effort that is ``\textit{demanding, especially when many observations are involved}''. Therefore, we developed this view based on juxtaposition.

\subsection{Ranking Projection Axis View}
\par The ranking projection axis view allows to compare ranking results from the ranking tabular view with the projection axis generated in the interactive projection view (\textbf{R.4}), which consists of a \textit{projection axis}, a \textit{score axis}, a \textit{contribution axis} and an \textit{attribute comparison} subviews.

\par \textbf{Projection Axis.} Following the generation of the project axis, we design the project axis subview (\autoref{fig:rankingprojectionaxis}(A)(1)). \quan{The black line} represents the turning point (the ``average observation'' of each rating) of the ranking line that decreases from left (high rating) to right (low rating). On the projection axis, the gray, blue, and orange red dots represent the observation. Their horizontal positions represent their projection location on the ranking line, and the height of the dots indicates the distance from the observation to the nearest ranking line. The gray color indicates the dots of which the ranking results are consistent with those in the projection, while blue and orange red indicate inconsistencies. For the projection axis, the distance to the 1D curve is computed, and this approach may place points that are on different sides of the curve to close or even identical positions. However, we did not choose a mapping to a signed distance from the curve to ensure that points that are left/right of the curve remain left/right. Borrowing the idea of an ``inverse ordinal number'', we further propose the concept of a positive/negative inverse ordinal number. For example, as shown in \autoref{fig:projectionaxis}(B), from $Rating1 > Rating2 > Rating3 > Rating4$ and $D2(Rating4) \in (Rating1, Rating2)$, we can ascertain that $Rating1 > D2(Rating4) > Rating2 > Rating3 > Rating4$. As two grades $Rating2$ and $Rating3$ are behind $D2(Rating4)$, $D2$ has a positive inverse ordinal value of $2$. Similarly, $D3$ has a negative inverse value of $1$. Blue indicates all positive inverse ordinal dots (the performance in the projection is better than that in the ranking), and red indicates the negative ones (the performance in the projection is worse than that in the ranking). The darker the color, the greater the values of the inverse ordinal.

\par \textbf{Score and Contribution Axis.} In the design of the score axis (\autoref{fig:rankingprojectionaxis}(A)(2)), each grey bar represents one item, and the items are ordered based on their ranking scores (i.e., the higher the score, the better). The height of the bars indicates the ranking scores. We render the score axis in an area chart to observe the changes in the ranking scores. In the design of the contribution axis (\autoref{fig:rankingprojectionaxis}(A)(3)), \quan{we use themeriver to represent the contribution of different data attributes to the ranking score of each item. The horizontal direction of the theme river design is consistent with that of the score axis. Analysts can observe how attributes' contributions change with the ranking score. Other designs like ordinary stacked bars could also be adopted but themeriver can distinguish different axes in different subviews.}

\par \textbf{Attribute Comparison.} In \autoref{fig:rankingprojectionaxis}(A)(4), each row represents one attribute and each column represents one item. The bar length of each column indicates the attribute value. In the initial state, the horizontal direction of the score axis, contribution axis, and the attribute comparison view have the same meaning, i.e., the same data item is in the same column. We use lines to connect the identical data items in the projection axis, score axis, contribution axis, and attribute comparison subviews. For example, clicking ``Huaxia Bank'' generates a line that connects all the identical items in the four subviews. As shown in \autoref{fig:rankingprojectionaxis}(A), when selecting ``Huaxia Bank'', the colors of all the bars indicate the differences of the attribute values between other banks and the selected bank. If the attribute value of the other bank is larger than that of the selected bank, the color of the bar of the other bank becomes blue. The larger the difference, the deeper the color. If the attribute value of the other bank is smaller than that of the selected bank, the color of the bar of the other bank becomes orange red and the smaller the difference, the deeper the color. The ``Align'' button enables bank classification according to the similarity between the selected and the other items (\autoref{fig:rankingprojectionaxis}(B)). The similarity is defined as the reciprocal of the square root sign of the attribute value difference between two items.

\subsection{Interactions Among the Views}
\par Rich interactions are integrated to catalyze an efficient in-depth analysis. \textit{(1) Zoom in/out.} \textit{RankAxis} leverages zoom in/out to facilitate inspections. As shown in \autoref{fig:teaser}(B), we can zoom in the area for a detailed observation of data attributes to resolve the visual clutter issue caused by the overlap of projection observations. \textit{(2) Click, Hover, and Link.} When hovering on elements on views like \autoref{fig:teaser}(B)(D)(E), detailed information of the elements is displayed as a tooltip. Users can click the align button in Ranking Projection Axis View, and the contribution comparison axis can switch the ranking order and similarity order. \textit{(3) Drag, Select, and Filter.} Analysts can adjust the ranking position of a particular item by dragging or customize a ranking line and filter items to observe their distributions in other views. For example, users can customize the ranking line and circle items to generate the rating line in the self-defined rating line mode.

\section{Evaluation}

\subsection{Case Study: Bank Rating}
\par
\begin{figure}[h]
  \vspace{-3mm}
 \centering 
 \includegraphics[width=\linewidth]{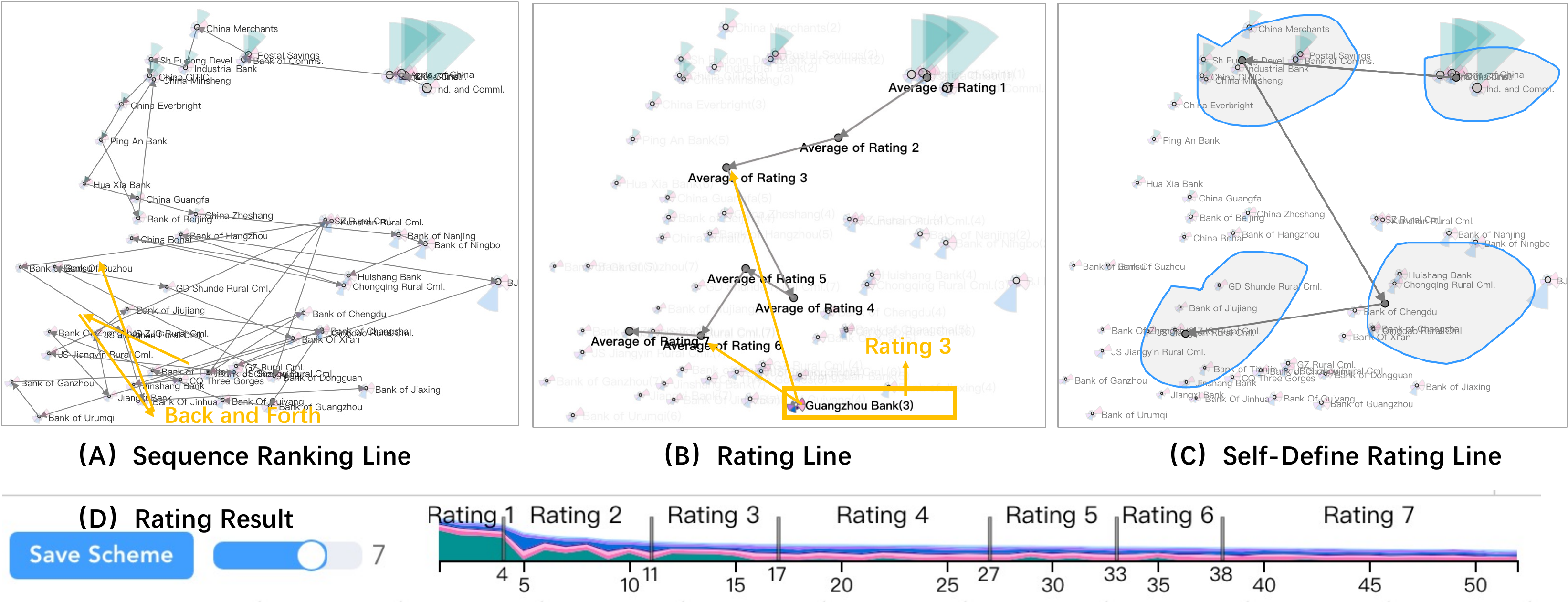}
    \vspace{-8mm}
 \caption{The experts leveraged \textit{RankAxis} to inspect the bank rating problem and identified some inconsistencies.}
 \label{fig:case1}
   \vspace{-3mm}
\end{figure}

 We show how E.1 explores inconsistencies in predictions and rankings by loading the 2018 Chinese bank credit rating data into \textit{RankAxis}. In \autoref{fig:case1}(A), he observed that banks are connected from high to low scores based on the ranking scores. He noted that the layout of the banks shows a specific trend regarding the direction, i.e., the ranking lines show a zigzag shape, while the projected layout generally conveys the semantic ranking direction but is not fully consistent with the ranking results. However, since all the banks are linked together, it was difficult for him to identify the exact direction of the ranking lines.

\par
\begin{figure}[h]
  \vspace{-3mm}
 \centering 
 \includegraphics[width=\linewidth]{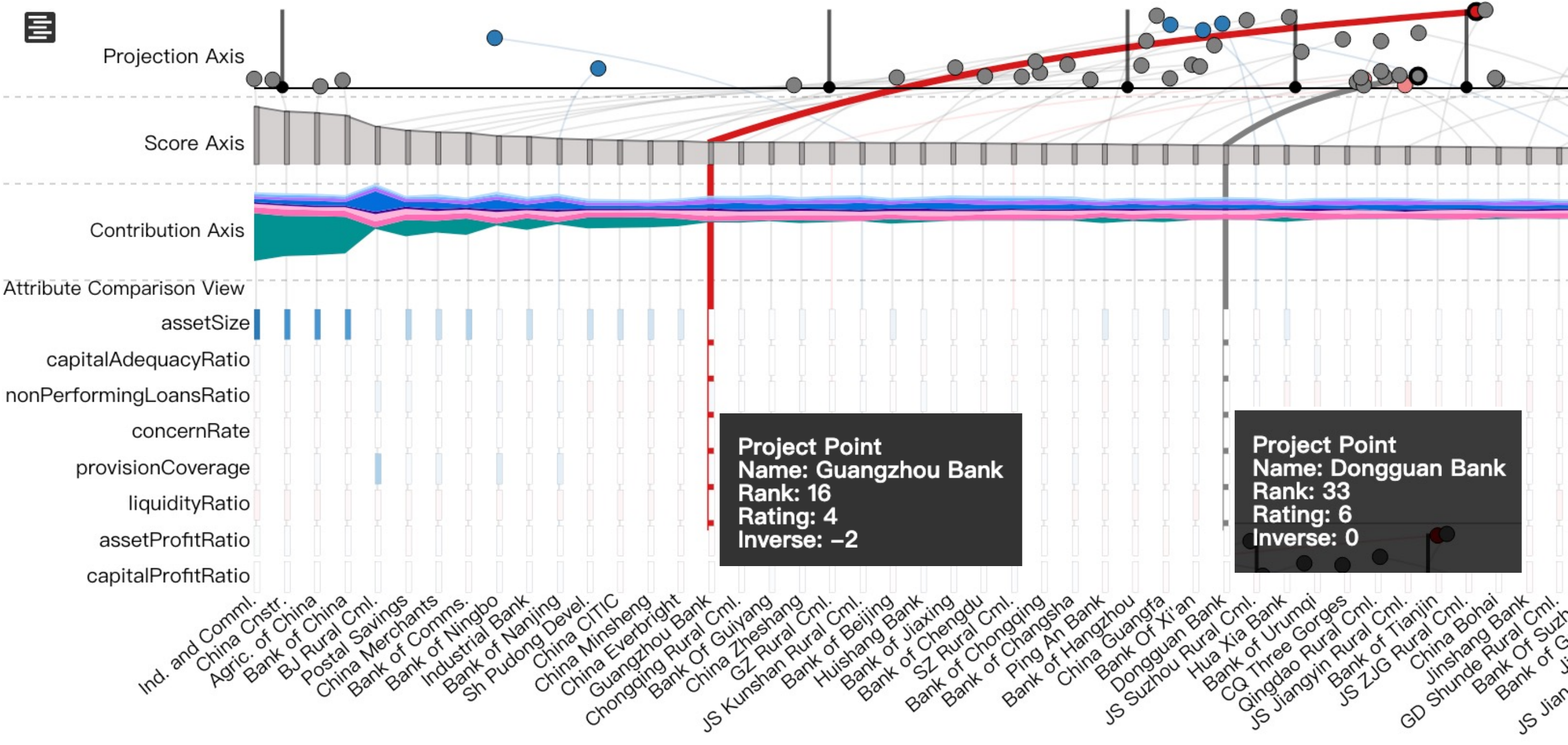}
   \vspace{-8mm}
 \caption{Guangzhou Bank is rated next to the average observation of rating 6 instead of rating 3.}
 \label{fig:case11}
     \vspace{-3mm}
\end{figure}

 To further check the rating issue, E.1 examined the ranking tabular view. He adjusted the slider to truncate the ranking list into seven ratings, as shown in \autoref{fig:case1}(D). \textit{RankAxis} automatically divides the ranking list according to the attribute contribution distribution, e.g., $Rating1$ for $1$ to $4$ and $Rating2$ for $5$ to $11$. Then, \textit{RankAxis} concatenates the ``average observations'' corresponding to each rating sequentially to form a rating line (\autoref{fig:case1}(B)). The information near each dot indicates the name of the bank and the number in parentheses denotes the bank's rating. Note that the each bank's rating is roughly next to its corresponding ``average observation'' and decreases in order along the rating line. The expert was pleased to note that the generated rating line eliminates some noise effects. He further noted that in \autoref{fig:case1}(B), \textit{Guangzhou Bank} has a rating of $3$, but its actual rating is closer to the ``average observation'' of a rating of $6$. That is, the distance from \textit{Guangzhou Bank} to the ``average observation'' with a rating of $6$ is shorter than the distance from the ``average observation'' with a rating of $3$. The expert was quite curious about this observation. He consulted the ranking projection axis view (\autoref{fig:case11}) and confirmed that \textit{Guangzhou Bank} (as shown by the orange red line and orange red dot) performs poorly in the projection given the orange-red color. The expert further noticed that \textit{Dongguan Bank} (gray line) has a rating of $6$, and the two banks are pretty similar in the attribute comparison view. ``\textit{That is why Guangzhou Bank's rating is next to the average observation of rating 6,}'' said E.1. The expert also found that neither the ranking nor the predicted results were in line with his expectations. For example, in \autoref{fig:case1}(B), the banks in the yellow box are the ones with ratings of $2$ and $3$; however, the ``average observations'' with ratings of $2$ and $3$ are far from the banks in the yellow box.

\par
\begin{figure}[h]
  \vspace{-4mm}
 \centering 
 \includegraphics[width=\linewidth]{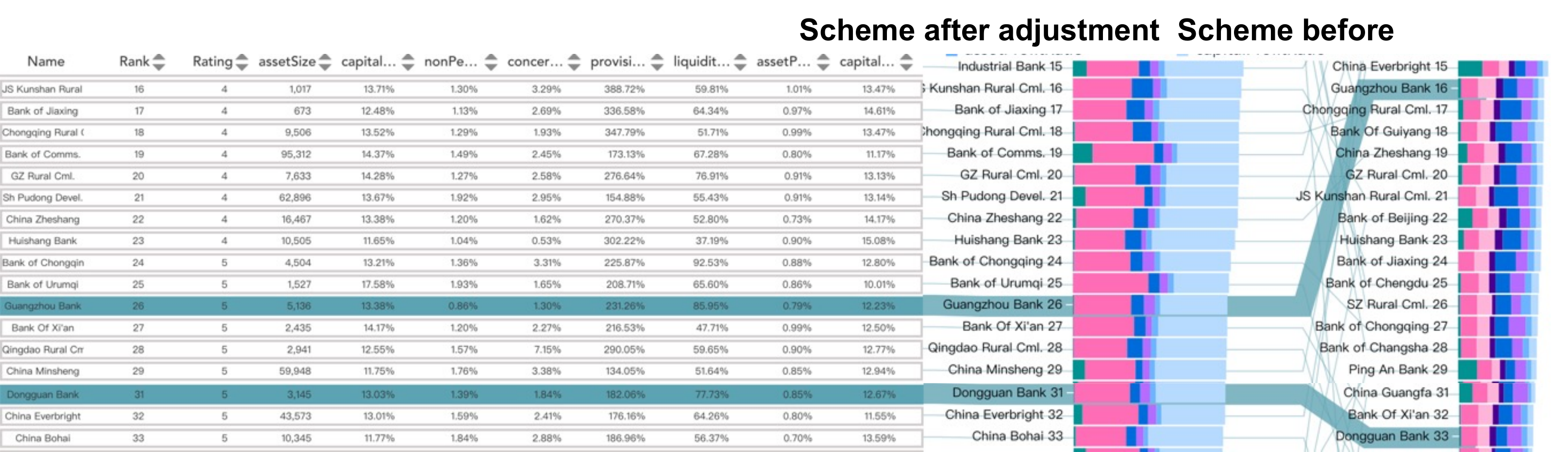}
   \vspace{-8mm}
 \caption{Adjustment Guangzhou Bank ranking to generating new ranking}
 \label{fig:case3}
     \vspace{-4mm}
\end{figure}

 E.1 thought Guangzhou Bank was overrate in ranking, so he compared \textit{Guangzhou Bank} and the neighboring banks and found that the assets of the neighboring banks are greater than \textit{Guangzhou Bank}, so E.1 wanted to put \textit{Guangzhou Bank} down in order, then he compared \textit{Guangzhou Bank} and the banks after it one by one to show as better than \textit{Jiangsu Kunshan Bank}, then adjusted \textit{Guangzhou Bank} from $16^{th}$ to $20^{th}$ and clicked ``Save Scheme''. After the adjustment, in \autoref{fig:case3}, Guangzhou Bank's ranking drops from $16^{th}$ to $26^{th}$, while \textit{Dongguan Bank}'s ranking rises from $33^{rd}$ to $31^{st }$. E.1 compared the rankings of the two banks with those of neighboring banks and indicated the results are reasonable. With this adjustment, the ranking distance between \textit{Guangzhou Bank} and \textit{Dongguan Bank} have become closer, and the consistency of rankings and projections has been improved.

\subsection{User Study}
\par \quan{In this subsection, we adopt a four-layer taxonomy~\cite{weibelzahl2020evaluation} and conduct a within-subjects study to evaluate \textit{RankAxis} in terms of \textit{informativeness}, \textit{effectiveness in decision-making}, \textit{usability} and \textit{visual design.}}

\par \quan{\textbf{Participants.} We recruit $18$ volunteers ($9$ females, $9$ males, age: $28 \pm 3.03$) for the user study. They are employees of the collaborated enterprise working in data analytics and machine learning. In particular, we select the participants with experience in data analysis and mining for bank ratings because they could provide us with more comprehensive insights and help us validate the usability of the system.}

\begin{figure}[h]
 \centering
   \vspace{-3mm}
 \includegraphics[width=\linewidth]{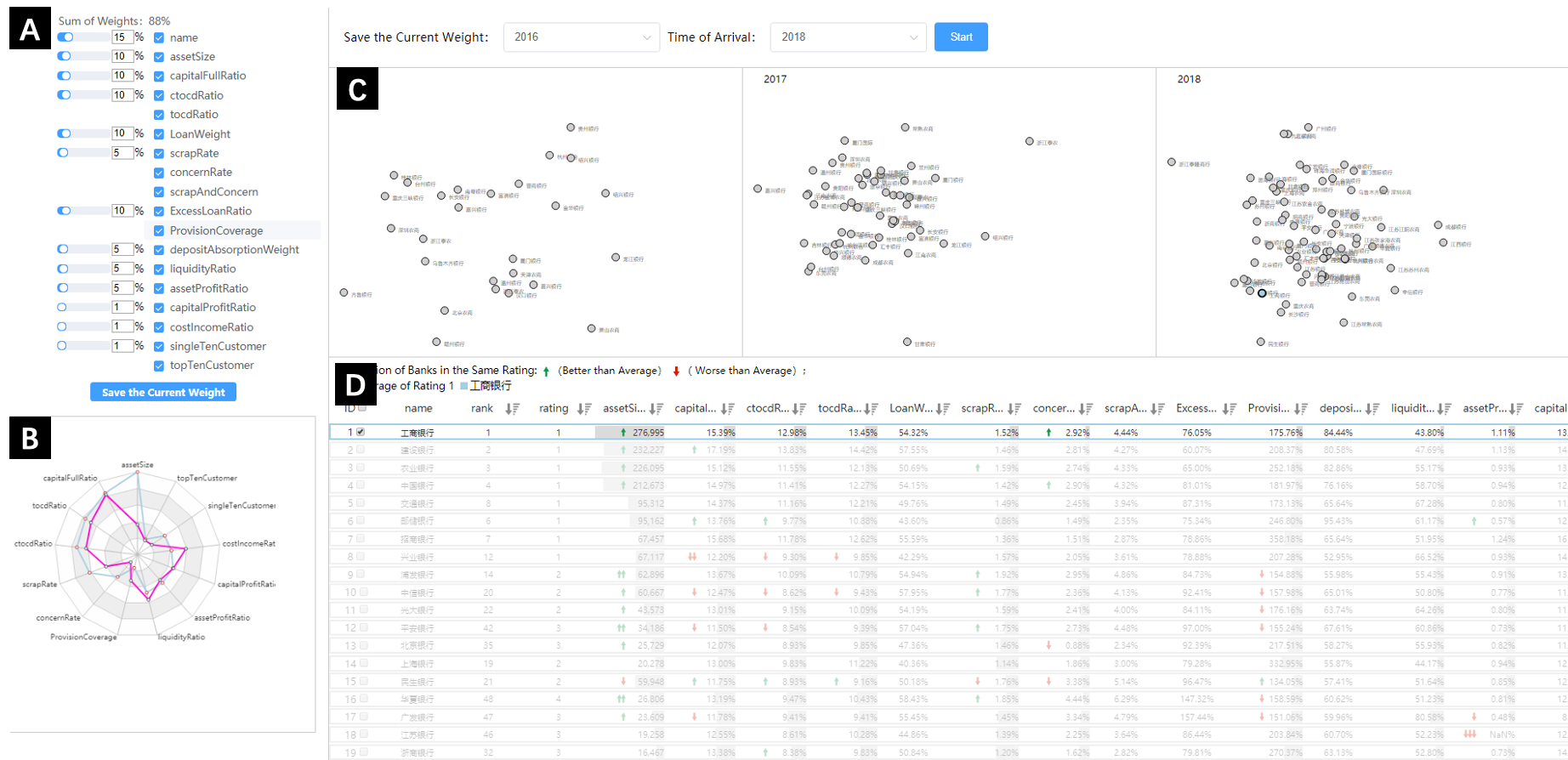}
 \vspace{-8mm}
 \caption{The primitive system: (A) Attribute weight adjusting panel supports users to specify. (B) Radar chart shows attribution value distribution for a certain data item. (C) Projection view shows 2D projection distributions corresponding to different weight settings. (D) Ranking tabular views show the ranking in the current weight setting.}
 \label{fig:primitive}
   \vspace{-3mm}
\end{figure}

\par \quan{\textbf{Experiment setup.} Since a formal comparative study with existing ranking or projection visualization systems is not applicable because previous work mainly focuses on one aspect and only covers a part of the previously mentioned requirements, we compare \textit{RankAxis} with one primitive system (\autoref{fig:primitive}). The difference between the primitive system and \textit{RankAxis} lie in: (1) The primitive system shares basic functions of the projection view and ranking tabular view with \textit{RankAxis}. However, the primitive system does not support interactively constructing the projection axis in the projection view, only showing the observation distribution without additional encodings. The ranking tabular view displays the data items' properties and rankings but does not support interactively adjustment the ranking by user interaction, e.g., drag-and-drop operation. (2) The primitive system requires a manual configuration on attribute weights by either specifying the value or adjusting a slider. Instead, \textit{RankAxis} supports direct manipulation on data items and the underlying algorithm infers the corresponding attribute weights. (3) \textit{RankAxis} employs more visual cues to support inferring inconsistency between ranking and projection while the primitive system does not support inconsistency inspection. To minimize the ordering and learning effect, we counterbalance the two systems.}

\par \quan{\textbf{Procedure.} We conduct the experiment in four sessions. In the first session, participants are briefed about the background, purpose and procedure of the experiment. Each following session last around $15$ minutes and one of the two systems is presented, briefed, and tested. Each participant is required to conduct two tasks with the provided system. The first task is to assign a reasonable rating to all the involved banks on the basis of the participants' domain knowledge. The second task is to evaluate any potential inconsistency in the bank rating between projection and ranking, such as an unreasonable ranking position of a certain bank in one method and explain the reason. Participants are allowed to think aloud their ideas when performing all the tasks. After finishing all the tasks with a particular system, they are required to complete a questionnaire with $7$-point Likert scale questions.}

\par \quan{We propose the following hypotheses: \textit{\textbf{H1.}} \textit{RankAxis} is more informative than the primitive system. Specifically, the information accessibility (\textit{H1a}), richness (\textit{H1b}), and sufficiency (\textit{H1c}) of \textit{RankAxis} is better than that of the primitive one. \textit{\textbf{H2.}} \textit{RankAxis} performs better than the primitive system in facilitating decision-making in terms of confidence (\textit{H2a}) and assistance (\textit{H2b}). \textit{\textbf{H3.}} The primitive system is preferred over the full one, i.e., more intuitive (\textit{H3a}), easier to comprehend (\textit{H3b}), learn (\textit{H3c}), and use (\textit{H3d}), and thus is better recommended (\textit{H3e}) than \textit{RankAxis}. We report the participants' quantitative ratings and verbal feedback on informativeness, decision-making efficacy, visual designs and usability, and run repeated measures ANOVA on each questionnaire item and the Bonferroni post-hoc test on measures with statistically significant differences. Upon completion of the user study, they will receive a $10$ coupon for an online e-commerce platform.}

\par \quan{\textbf{Informativeness and decision-making efficacy.} \textit{RankAxis} receives significantly higher scores in almost all the studied metrics of informativeness except the metric of information accessibility than the primitive system. Participants find assessing information is significantly easier in the primitive system than \textit{RankAxis} (H1a rejected). However, regarding information richness and sufficiency, there exists a significant difference between the two systems, i.e., the information offered by \textit{RankAxis} is shown to be richer and more sufficient in rating banks compared with the primitive system. ``\textit{With RankAxis, I can obtain some insights that I might never be able to notice before.}'' (P12, male, age: $33$) As such, participants report significantly higher confidence in rating banks and identifying the inconsistencies between projection and ranking results using \textit{RankAxis} compared with the primitive system (H2a supported). Participants also report that \textit{RankAxis} provides significantly more assistance than the baseline (H2b supported). In summary, the results on informativeness and decision-making efficacy demonstrate that \textit{RankAxis} provides more rich and sufficient information, although less accessible than the primitive system.}

\par \quan{\textbf{Intuitiveness and comprehension.} Different from our hypothesis, the primitive system is not significantly more intuitive or comprehensive than \textit{RankAxis}. Participants report that they can easily manipulate items to adjust rankings which is ``\textit{far more intuitive than filling numbers on attribute weights.}'' (P6, male, age: $26$) Participants report that although the ranking projection axis view is not very intuitive, it carries more information that the primitive system fails to convey.}

\par \quan{\textbf{Learn, use and recommendable.} We do not notice a significant difference regarding easy to learn and use among the two systems ($p=.15$) (H3c, H3d rejected). ``\textit{They all need a learning curve.}'' (P2, male, age: $28$) Also, no significant difference is found regarding recommendation between the two systems ($p=.35$, H3e rejected). ``\textit{When I just need a quick glance at the banks' performance, I will recommend the primitive system; however, I will turn to RankAxis if I need to dig out some deeper insights.}'' (P15, female, age: $27$)}

\subsection{Expert Interview}
\par We conducted $30$-minute semi-structured interviews with the team of credit rating experts (E.1 -- E.4). They all appreciated the capability of \textit{RankAxis} to support interactive exploration of projection and ranking results. E.1 remarked that \textit{RankAxis} has greatly improved productivity by allowing him to easily obtain and interpret projection and rating results in the same context. They were pleased with the flexible interaction and visualization, as it allowed ``\textit{to effectively raise potential inconsistencies.}'' We conducted a user-centered design process inspired by the experts' routines to inform the system design, such as the inclusion of a table and projection view of the rankings. After a brief introduction, the experts developed a customized exploration path. The experts also noted that our design is quite generic, as it is already suitable for other applications. For example, the projection and ranking tabular views can be applied to almost any rating problem. Regarding scalability, we used more than $10$ colors to distinguish attributes. However, we are fully aware that only a few colors can be effectively used as category labels~\cite{2012Information}. When the number of attributes exceeds $12$, a design like the stacked graph is not appropriate, and for scenarios with more attributes, filtering should be supported to display the attributes of interest.

\par Regarding to comparison with traditional practices using \textit{Excel} and \textit{Tableau}, participants reported that they are both powerful tools for ranking analysis. However, they all agreed that \textit{Excel} or \textit{Tableau} only support a small part of the requirement analysis. We believe this is because these tools and \textit{RankAxis} follow different decision paths: separate ranking analysis and joint analysis of projection and ranking: ``\textit{tools such as \textit{Tableau} provide only a ranking overview, and we have iterate multiple ranking interactions to determine the strong and weak attributes of different banks}'' (P12, female, age: $33$). Furthermore, participants found that the weighting process in \textit{Tableau} or \textit{Excel} suspicious, and they usually did not know if they had achieved the appropriate weighting, in line with their expectations. In short, users need to use \textit{Excel} and \textit{Tableau} when they know their dataset well and do not need to interpret the results. \textit{RankAxis} complements their tools to explore the projection and ranking results together.

\section{Discussion and Limitation}
\par \textbf{Contributions over previous work.} Projection is ubiquitous in many visualization systems because of its ability to translate high-dimensional planes into low dimensions, but this planarization has limitations and users may be misled by this representation. Ranking, also ubiquitous, may also be affected by differences in its interpretation of the relative importance of features in ranking. Very little work intends to overcome these limitations and point out potential connections between them at the same time. We combine projection with ranking in order to combine strengths and reduce weaknesses. \quan{This work has the potential to be applied to a variety of usage scenarios such as using multivariate university ranking data to find dream graduate schools.}

\par \textbf{Choice of dimensionality reduction algorithms.} \textit{RankAxis} is independent of the projection method, as long as it can reveal the potential global and/or local structure of the high-dimensional data of interest. We should also note that when using \textit{RankAxis} to analyze linear or nonlinear semantics, the rating results can be either accurate or approximate. For example, for linear dimensionality reduction algorithms such as \textit{PCA}, any direction in the reduced space can be represented as a linear combination of the original dimensions, which can be directly and accurately revealed by \textit{RankAxis}; for nonlinear dimensionality reduction algorithms such as \textit{t-SNE} and \textit{UMAP}~\cite{Mcinnes2018UMAP}, \textit{RankAxis} can provide a linear approximation to the nonlinear semantics they reveal.

\par \textbf{Limitation.} \quan{First, \textit{RankAxis} is currently tailored to investigate rankings while conducting segmentation for ratings.} Second, we do not support non-numeric attributes. For numeric attributes, only maximum values (the larger the better) and minimum values (the smaller the better) are supported. In other words, intermediate values (closer to the middle value is better) and interval values (within a certain range is better) are not supported. Third, despite the application of semantic zooming, visual clutter can occur in the projection view. Fourth, we use \textit{Ranking SVM} to infer weights for changes based on a subset of items; and this approach may occasionally perform poorly because the weights only represent local changes.

\section{Conclusion and Future Work}
\par We present \textit{RankAxis}, a visual analytics system that combines projection and ranking into the same exploration environment and facilitates the mutual interpretation of their results. A case study and a user study validate the efficacy of \textit{RankAxis}. We will further consider cases with more different types of attributes and develop more complex classifiers that assign different sets of weights to different types of data items.

\acknowledgments{
We are grateful for the valuable feedback and comments provided by the anonymous reviewers. This work is partially supported by the Research Start-up Fund of ShanghaiTech University and the Research Grants Council of the Hong Kong Special Administrative Region, China under General Research Fund (GRF) with Grant No.: 16204420.}

\bibliographystyle{abbrv-doi}

\balance
\bibliography{template}

\begin{thebibliography}{10}

\bibitem{albuquerque2010improving}
G.~Albuquerque, M.~Eisemann, D.~J. Lehmann, H.~Theisel, and M.~Magnor.
\newblock Improving the visual analysis of high-dimensional datasets using
  quality measures.
\newblock In {\em 2010 IEEE Symposium on Visual Analytics Science and
  Technology}, pp. 19--26. IEEE, 2010.

\bibitem{behrisch2018quality}
M.~Behrisch, M.~Blumenschein, N.~W. Kim, L.~Shao, M.~El-Assady, J.~Fuchs,
  D.~Seebacher, A.~Diehl, U.~Brandes, H.~Pfister, et~al.
\newblock Quality metrics for information visualization.
\newblock In {\em Computer Graphics Forum}, vol.~37, pp. 625--662. Wiley Online
  Library, 2018.

\bibitem{blumenschein2018smartexplore}
M.~Blumenschein, M.~Behrisch, S.~Schmid, S.~Butscher, D.~R. Wahl, K.~Villinger,
  B.~Renner, H.~Reiterer, and D.~A. Keim.
\newblock Smartexplore: Simplifying high-dimensional data analysis through a
  table-based visual analytics approach.
\newblock In {\em 2018 IEEE Conference on Visual Analytics Science and
  Technology (VAST)}, pp. 36--47. IEEE, 2018.

\bibitem{carenini2004valuecharts}
G.~Carenini and J.~Loyd.
\newblock Valuecharts: analyzing linear models expressing preferences and
  evaluations.
\newblock In {\em Proceedings of the working conference on Advanced visual
  interfaces}, pp. 150--157, 2004.

\bibitem{coimbra2021analyzing}
D.~B. Coimbra, R.~M. Martins, E.~Mota, T.~Tiburtino, P.~Diamantino, and M.~L.
  Peixoto.
\newblock Analyzing the quality of local and global multidimensional
  projections using performance evaluation planning.
\newblock {\em Theoretical Computer Science}, 2021.

\bibitem{dasgupta2010pargnostics}
A.~Dasgupta and R.~Kosara.
\newblock Pargnostics: Screen-space metrics for parallel coordinates.
\newblock {\em IEEE Transactions on Visualization and Computer Graphics},
  16(6):1017--1026, 2010.

\bibitem{donoho2000high}
D.~L. Donoho et~al.
\newblock High-dimensional data analysis: The curses and blessings of
  dimensionality.
\newblock {\em AMS math challenges lecture}, 1(2000):32, 2000.

\bibitem{ellis2006plot}
G.~Ellis and A.~Dix.
\newblock The plot, the clutter, the sampling and its lens: occlusion measures
  for automatic clutter reduction.
\newblock In {\em Proceedings of the working conference on Advanced visual
  interfaces}, pp. 266--269, 2006.

\bibitem{espadoto2019towards}
M.~Espadoto, R.~M. Martins, A.~Kerren, N.~S. Hirata, and A.~C. Telea.
\newblock Towards a quantitative survey of dimension reduction techniques.
\newblock {\em IEEE Transactions on Visualization and Computer Graphics}, 2019.

\bibitem{fayyad1993multi}
U.~Fayyad and K.~Irani.
\newblock Multi-interval discretization of continuous-valued attributes for
  classification learning.
\newblock 1993.

\bibitem{fernstad2013quality}
S.~J. Fernstad, J.~Shaw, and J.~Johansson.
\newblock Quality-based guidance for exploratory dimensionality reduction.
\newblock {\em Information Visualization}, 12(1):44--64, 2013.

\bibitem{fewdesigning}
S.~Few and P.~E. Principal.
\newblock Designing tables and graphs to enlighten.

\bibitem{gleicher2013explainers}
M.~Gleicher.
\newblock Explainers: Expert explorations with crafted projections.
\newblock {\em IEEE transactions on visualization and computer graphics},
  19(12):2042--2051, 2013.

\bibitem{gratzl2013lineup}
S.~Gratzl, A.~Lex, N.~Gehlenborg, H.~Pfister, and M.~Streit.
\newblock Lineup: Visual analysis of multi-attribute rankings.
\newblock {\em IEEE Transactions on Visualization and Computer Graphics},
  19(12):2277--2286, 2013.

\bibitem{grinstein2002information}
U.~M. F. G.~G. Grinstein and A.~Wierse.
\newblock {\em Information visualization in data mining and knowledge
  discovery}.
\newblock Morgan Kaufmann, 2002.

\bibitem{2018Interactive}
F.~Heimerl and M.~Gleicher.
\newblock Interactive analysis of word vector embeddings.
\newblock {\em Computer Graphics Forum}, 37(3):253--265, 2018.

\bibitem{inselberg1985plane}
A.~Inselberg.
\newblock The plane with parallel coordinates.
\newblock {\em The visual computer}, 1(2):69--91, 1985.

\bibitem{inselberg1990parallel}
A.~Inselberg and B.~Dimsdale.
\newblock Parallel coordinates: a tool for visualizing multi-dimensional
  geometry.
\newblock In {\em Proceedings of the First IEEE Conference on Visualization:
  Visualization90}, pp. 361--378. IEEE, 1990.

\bibitem{joachims2002optimizing}
T.~Joachims.
\newblock Optimizing search engines using clickthrough data.
\newblock In {\em Proceedings of the eighth ACM SIGKDD international conference
  on Knowledge discovery and data mining}, pp. 133--142, 2002.

\bibitem{kammer2020glyphboard}
D.~Kammer, M.~Keck, T.~Gr{\"u}nder, A.~Maasch, T.~Thom, M.~Kleinsteuber, and
  R.~Groh.
\newblock Glyphboard: Visual exploration of high-dimensional data combining
  glyphs with dimensionality reduction.
\newblock {\em IEEE Transactions on Visualization and Computer Graphics},
  26(4):1661--1671, 2020.

\bibitem{kim2015interaxis}
H.~Kim, J.~Choo, H.~Park, and A.~Endert.
\newblock Interaxis: Steering scatterplot axes via observation-level
  interaction.
\newblock {\em IEEE transactions on visualization and computer graphics},
  22(1):131--140, 2015.

\bibitem{kwon2016axisketcher}
B.~C. Kwon, H.~Kim, E.~Wall, J.~Choo, H.~Park, and A.~Endert.
\newblock Axisketcher: Interactive nonlinear axis mapping of visualizations
  through user drawings.
\newblock {\em IEEE Transactions on Visualization and Computer Graphics},
  23(1):221--230, 2016.

\bibitem{lai1994topsis}
Y.-J. Lai, T.-Y. Liu, and C.-L. Hwang.
\newblock Topsis for modm.
\newblock {\em European journal of operational research}, 76(3):486--500, 1994.

\bibitem{li2018embeddingvis}
Q.~Li, K.~S. Njotoprawiro, H.~Haleem, Q.~Chen, C.~Yi, and X.~Ma.
\newblock Embeddingvis: A visual analytics approach to comparative network
  embedding inspection.
\newblock In {\em 2018 IEEE Conference on Visual Analytics Science and
  Technology (VAST)}, pp. 48--59. IEEE, 2018.

\bibitem{li2021semanticaxis}
Z.~Li, C.~Zhang, Y.~Zhang, and J.~Zhang.
\newblock Semanticaxis: exploring multi-attribute data by semantic construction
  and ranking analysis.
\newblock {\em Journal of Visualization}, 24(5):1065--1081, 2021.

\bibitem{liu2021inspecting}
Q.~Liu, Q.~Li, Z.~Zhu, T.~Ye, and X.~Ma.
\newblock Inspecting the process of bank credit rating via visual analytics.
\newblock In {\em 2021 IEEE Visualization Conference (VIS)}, pp. 136--140.
  IEEE, 2021.

\bibitem{liu2016visualizing}
S.~Liu, D.~Maljovec, B.~Wang, P.-T. Bremer, and V.~Pascucci.
\newblock Visualizing high-dimensional data: Advances in the past decade.
\newblock {\em IEEE Transactions on Visualization and Computer Graphics},
  23(3):1249--1268, 2016.

\bibitem{maaten2008visualizing}
L.~v.~d. Maaten and G.~Hinton.
\newblock Visualizing data using t-sne.
\newblock {\em Journal of machine learning research}, 9(Nov):2579--2605, 2008.

\bibitem{mcdonnell2008illustrative}
K.~T. McDonnell and K.~Mueller.
\newblock Illustrative parallel coordinates.
\newblock In {\em Computer Graphics Forum}, vol.~27, pp. 1031--1038. Wiley
  Online Library, 2008.

\bibitem{Mcinnes2018UMAP}
L.~Mcinnes and J.~Healy.
\newblock Umap: Uniform manifold approximation and projection for dimension
  reduction.
\newblock {\em Journal of Open Source Software}, 3(29):861, 2018.

\bibitem{munzner2014visualization}
T.~Munzner.
\newblock {\em Visualization analysis and design}.
\newblock CRC press, 2014.

\bibitem{nonato2018multidimensional}
L.~G. Nonato and M.~Aupetit.
\newblock Multidimensional projection for visual analytics: Linking techniques
  with distortions, tasks, and layout enrichment.
\newblock {\em IEEE Transactions on Visualization and Computer Graphics},
  25(8):2650--2673, 2018.

\bibitem{pajer2016weightlifter}
S.~Pajer, M.~Streit, T.~Torsney-Weir, F.~Spechtenhauser, T.~M{\"o}ller, and
  H.~Piringer.
\newblock Weightlifter: Visual weight space exploration for multi-criteria
  decision making.
\newblock {\em IEEE transactions on visualization and computer graphics},
  23(1):611--620, 2016.

\bibitem{peng2004clutter}
W.~Peng, M.~O. Ward, and E.~A. Rundensteiner.
\newblock Clutter reduction in multi-dimensional data visualization using
  dimension reordering.
\newblock In {\em IEEE Symposium on Information Visualization}, pp. 89--96.
  IEEE, 2004.

\bibitem{puri2020rankbooster}
A.~Puri, B.~K. Ku, Y.~Wang, and H.~Qu.
\newblock Rankbooster: Visual analysis of ranking predictions.
\newblock {\em arXiv preprint arXiv:2004.06435}, 2020.

\bibitem{rao1994table}
R.~Rao and S.~K. Card.
\newblock The table lens: merging graphical and symbolic representations in an
  interactive focus+ context visualization for tabular information.
\newblock In {\em Proceedings of the SIGCHI conference on Human factors in
  computing systems}, pp. 318--322, 1994.

\bibitem{seo2004rank}
J.~Seo and B.~Shneiderman.
\newblock A rank-by-feature framework for unsupervised multidimensional data
  exploration using low dimensional projections.
\newblock In {\em IEEE Symposium on Information Visualization}, pp. 65--72.
  IEEE, 2004.

\bibitem{sorzano2014survey}
C.~O.~S. Sorzano, J.~Vargas, and A.~P. Montano.
\newblock A survey of dimensionality reduction techniques.
\newblock {\em arXiv preprint arXiv:1403.2877}, 2014.

\bibitem{tatu2009combining}
A.~Tatu, G.~Albuquerque, M.~Eisemann, J.~Schneidewind, H.~Theisel, M.~Magnork,
  and D.~Keim.
\newblock Combining automated analysis and visualization techniques for
  effective exploration of high-dimensional data.
\newblock In {\em 2009 IEEE Symposium on Visual Analytics Science and
  Technology}, pp. 59--66. IEEE, 2009.

\bibitem{tufte1985visual}
E.~R. Tufte.
\newblock The visual display of quantitative information.
\newblock {\em The Journal for Healthcare Quality (JHQ)}, 7(3):15, 1985.

\bibitem{tufte1990envisioning}
E.~R. Tufte, N.~H. Goeler, and R.~Benson.
\newblock {\em Envisioning information}, vol. 126.
\newblock Graphics press Cheshire, CT, 1990.

\bibitem{tzeng2011multiple}
G.-H. Tzeng and J.-J. Huang.
\newblock {\em Multiple attribute decision making: methods and applications}.
\newblock CRC press, 2011.

\bibitem{van2009dimensionality}
L.~Van Der~Maaten, E.~Postma, and J.~Van~den Herik.
\newblock Dimensionality reduction: a comparative.
\newblock {\em J Mach Learn Res}, 10(66-71):13, 2009.

\bibitem{wall2017podium}
E.~Wall, S.~Das, R.~Chawla, B.~Kalidindi, E.~T. Brown, and A.~Endert.
\newblock Podium: Ranking data using mixed-initiative visual analytics.
\newblock {\em IEEE Transactions on Visualization and Computer Graphics},
  24(1):288--297, 2017.

\bibitem{2012Information}
C.~Ware.
\newblock {\em Information Visualization: Perception for Design}.
\newblock Morgan Kaufmann Publishers Inc., 2012.

\bibitem{weibelzahl2020evaluation}
S.~Weibelzahl, A.~Paramythis, and J.~Masthoff.
\newblock Evaluation of adaptive systems.
\newblock In {\em Proceedings of the 28th ACM Conference on User Modeling,
  Adaptation and personalization}, pp. 394--395, 2020.

\bibitem{weng2018srvis}
D.~Weng, R.~Chen, Z.~Deng, F.~Wu, J.~Chen, and Y.~Wu.
\newblock Srvis: Towards better spatial integration in ranking visualization.
\newblock {\em IEEE Transactions on Visualization and Computer Graphics},
  25(1):459--469, 2018.

\bibitem{wenskovitch2017towards}
J.~Wenskovitch, I.~Crandell, N.~Ramakrishnan, L.~House, and C.~North.
\newblock Towards a systematic combination of dimension reduction and
  clustering in visual analytics.
\newblock {\em IEEE Transactions on Visualization and Computer Graphics},
  24(1):131--141, 2017.

\bibitem{xia2017ldsscanner}
J.~Xia, F.~Ye, W.~Chen, Y.~Wang, W.~Chen, Y.~Ma, and A.~K. Tung.
\newblock Ldsscanner: Exploratory analysis of low-dimensional structures in
  high-dimensional datasets.
\newblock {\em IEEE Transactions on Visualization and Computer Graphics},
  24(1):236--245, 2017.

\bibitem{yates2014visualizing}
A.~Yates, A.~Webb, M.~Sharpnack, H.~Chamberlin, K.~Huang, and R.~Machiraju.
\newblock Visualizing multidimensional data with glyph sploms.
\newblock In {\em Computer Graphics Forum}, vol.~33, pp. 301--310. Wiley Online
  Library, 2014.

\bibitem{yuan2013dimension}
X.~Yuan, D.~Ren, Z.~Wang, and C.~Guo.
\newblock Dimension projection matrix/tree: Interactive subspace visual
  exploration and analysis of high dimensional data.
\newblock {\em IEEE Transactions on Visualization and Computer Graphics},
  19(12):2625--2633, 2013.

\bibitem{zhao2017skylens}
X.~Zhao, Y.~Wu, W.~Cui, X.~Du, Y.~Chen, Y.~Wang, D.~L. Lee, and H.~Qu.
\newblock Skylens: Visual analysis of skyline on multi-dimensional data.
\newblock {\em IEEE Transactions on Visualization and Computer Graphics},
  24(1):246--255, 2017.

\end{thebibliography}
\end{document}